\documentclass[modern, astrosymb]{aastex631}

\usepackage[activate={true,nocompatibility},final,tracking=true,kerning=true]{microtype}
\usepackage{chemformula}
\usepackage{csquotes}
\let\tablenum\relax
\usepackage{siunitx}
\usepackage{booktabs}
\usepackage{multirow, booktabs}
\usepackage{lipsum}
\usepackage{lmodern}
\usepackage{tcolorbox}
\usepackage{graphicx}
\usepackage{appendix}
\usepackage{changepage}
\makeatletter
\let\frontmatter@title@above=\relax
\makeatother

\usepackage[section]{placeins}

\definecolor{SOF}{rgb}{0.5, 0.0, 0.5}

\definecolor{NTT}{rgb}{0.0, 0.5, 0.5}

\definecolor{Jason}{rgb}{0.5, 0.5, 0.0}

\definecolor{EDIT}{rgb}{0.0, 0.5, 0.25}

\shorttitle{An ATA SETI Search of TRAPPIST-1}

\graphicspath{{./}{figures/}}

\begin{document}

\title{A Radio Technosignature Search of TRAPPIST-1 with the Allen Telescope Array}

\newcommand{\PSUAA}{Department of Astronomy \& Astrophysics, The Pennsylvania State University}
\newcommand{\CEHW}{Center for Exoplanets and Habitable Worlds}
\newcommand{\PSETI}{Penn State Extraterrestrial Intelligence Center}
\newcommand{\BL}{Breakthrough Listen, University of California, Berkeley}
\newcommand{\UCR}{Department of Earth \& Planetary Sciences, University of California, Riverside}
\newcommand{\UCB}{Department of Astronomy, University of California, Berkeley}
\newcommand{\SETI}{SETI Institute}

\correspondingauthor{Nick Tusay}
\email{tusay@psu.edu}

\author[0000-0001-9686-5890]{Nick Tusay}
\affiliation{\PSUAA}
\affiliation{\CEHW}
\affiliation{\PSETI}

\author[0000-0001-7057-4999]{Sofia Z. Sheikh}
\affiliation{\PSETI}
\affiliation{\BL}
\affiliation{\SETI}

\author[0000-0001-5290-1001]{Evan L. Sneed}
\affiliation{\UCR}
\affiliation{\BL}
\affiliation{\PSETI}

\author[0000-0002-0161-7243]{Wael Farah}
\affiliation{\BL}
\affiliation{\SETI}

\author[0000-0002-3430-7671]{Alexander W. Pollak}
\affiliation{\SETI}

\author[0000-0001-5576-2254]{Luigi F. Cruz}
\affiliation{\SETI}

\author[0000-0003-2828-7720]{Andrew Siemion}
\affiliation{\BL}
\affiliation{\SETI}
\affiliation{\UCB}

\author[0000-0003-3197-2294]{David R. DeBoer}
\affiliation{\UCB}

\author[0000-0001-6160-5888]{Jason T. Wright}
\affiliation{\PSUAA}
\affiliation{\CEHW}
\affiliation{\PSETI}


\begin{abstract} 

Planet-planet occultations (PPOs) occur when one exoplanet occults another exoplanet in the same system as seen from the Earth's vantage point. PPOs may provide a unique opportunity to observe radio ``spillover'' from extraterrestrial intelligences' (ETIs) radio transmissions or radar being transmitted from the further exoplanet towards the nearer one for the purposes of communication or scientific exploration. Planetary systems with many tightly packed, low-inclination planets, such as TRAPPIST-1, are predicted to have frequent PPOs. Here, the narrowband technosignature search code \texttt{turboSETI} was used in combination with the newly developed \texttt{NbeamAnalysis} filtering pipeline to analyze 28 hours of beamformed data taken with the Allen Telescope Array (ATA) during late October and early November 2022, from 0.9--9.3~GHz, targeting TRAPPIST-1. During this observing window, 7 possible PPO events were predicted using the \texttt{NbodyGradient} code. The filtering pipeline reduced the original list of 25 million candidate signals down to 6 million by rejecting signals that were not sky-localized and, from these, identified a final list of 11127 candidate signals above a power law cutoff designed to segregate signals by their attenuation and morphological similarity between beams. All signals were plotted for visual inspection, 2264 of which were found to occur during PPO windows. We report no detection of signals of non-human origin, with upper limits calculated for each PPO event exceeding EIRPs of 2.17--13.3 TW for minimally drifting signals and 40.8--421 TW in the maximally drifting case. This work constitutes the longest single-target radio SETI search of TRAPPIST-1 to date. 
\end{abstract}


\vspace{1cm}
\section{Introduction}
\label{sec:intro}
\medskip

Modern Searches for Extraterrestrial Intelligence (SETI) tend to focus on trying to detect unnaturally narrowband radio transmission. In natural radio science contexts, this can mean signals over frequency ranges less than $\sim$100 kHz, but in a SETI context narrowband usually means signals with widths measured in Hz \citep[n.b.\ some SETI work does focus on broader signals, e.g.][and the tools we use here are sensitive to these signals as well]{Broadband_Neural_Networks_Gajjar:2022:, Broadband_blips_Suresh:2023:}. While some terrestrial digital communication applications often opt for the higher data transmission rates of broadband modulation schemes, narrowband signals are useful as carrier waves for low signal-to-noise (SNR) cases, making it easy to find and lock onto a signal Doppler-drifting through frequency space, particularly in the case of our deep space probes like Voyager \citep{Voyager_SETI_Derrick:2023:}. In the limits of our sensitivity to interstellar radio transmission, for a given transmitter power, such narrowband signals offer the best possibility of detection.

Sufficiently powerful narrowband signals also have the potential to address ambiguous interpretations. Humans have developed technology to produce much narrower radio transmissions ($\lesssim$1~Hz) than natural astrophysical processes, suggesting that non-human sources of such signals must be artificially constructed by some biologically derived entity. The search for radio technosignatures is motivated by the fact that modern radio receivers on Earth can detect signals produced by the largest modern-day transmitters over interstellar distances \citep{Cocconi1959}. However, disentangling such distantly sourced signals from signals of anthropogenic origin can be tricky, particularly as more of the radio spectrum is being utilized for various terrestrial communication purposes. Ever since the very first Search for Extraterrestrial Intelligence (SETI) project, parsing out radio frequency interference (RFI) has been a significant problem \citep{Ozma_Drake:1961:}. Indeed, a major component of most modern radio SETI projects include both observational and data analysis strategies for RFI mitigation \citep{SETI_Tarter:2001:}. 

The standard observational strategy for single-dish radio SETI is to employ an on/off nodding cadence, in an attempt to discriminate persistent signals localized to a particular source \citep{nodding_Lebofsky:2019:}. However, radio arrays such as the Allen Telescope Array (ATA) employ interferometric techniques to combine the data from each dish in a wide field of view. Beamforming ---the interferometric addition of signals from many elements to form tight synthetic beams within the primary field of view--- can be used to synthesize many simultaneous, smaller and spatially separated beams, which provides a similar effect to the on/off nodding cadence without having to slew away from the target. This is the primary observational strategy employed at the ATA to distinguish RFI from sky-localized astrophysical signals \citep{ATA_beams_Harp:2005:}. 

The newly upgraded ATA was originally designed and recently upgraded with SETI in mind as the primary scientific driver \citep{refurbished_ATA_Sheikh:2023:}. The main focus of the research presented in this paper was to build a generalized pipeline for analyzing the SETI data products at the ATA and then apply it to a specific SETI observation. 

The TRAPPIST-1 system was selected as the primary science target for observation. Since the publication of its discovery in 2016 \citep{2016Natur.533..221G}, TRAPPIST-1 has provided an intriguing case study for exoplanet science. The star is a cool M-dwarf---the most abundant spectral type of star---and currently one of the best prospects for detecting and characterizing orbiting terrestrial-sized planets with solid surfaces due to the relative masses and radii between the star and these kinds of planets. Such stars, particularly later M-dwarfs, have very low luminosities that are difficult to observe at large distances. But the  proximity of TRAPPIST-1, at $\sim$12.5 parsecs, allows extremely precise measurements of the system's characteristics \citep{agolRefiningTransittimingPhotometric2021}. The TRAPPIST-1 system is almost perfectly edge-on from our line of sight, hosting seven known transiting planets. The planets are evenly spaced and in orbital resonance, with planets d, e, f and g spanning the extreme limits where liquid water could theoretically exist on a planetary surface given stellar irradiance and a sufficiently thick atmosphere \citep{Exoplanet_HZ_Catalog_Hill:2023:}. Much work has been done to determine the system's viability for biological development \citep{TRAPPIST_atmospheres_Turbet:2020:, TRAPPIST_Volatiles_Krissansen-Totton:2023:, TRAPPIST-1_Gillon:2024:}. 

The precisely measured orbital characteristics of this compact multi-planet system suggest that the system is so close to edge-on that several planets may occult each other in what is called a planet-planet occultation (PPO). There are many nearby multi-planet nearly co-planar systems that may exhibit PPOs, but the TRAPPIST-1 system is unique among these in how well constrained its many planets' orbital parameters are and how closely aligned they all are. The TRAPPIST-1 system is the perfect laboratory to both predict and potentially observe PPOs. Photometrically observing PPOs outside of transit is at the cusp of our technological capability, and it could be a useful observational tool in the near future \citep{lugerPlanetPlanetOccultations2017}. 

Even without being able to directly observe PPOs, these events may be critical observation windows for SETI if accurately predicted. On Earth, the Deep Space Network (DSN) is frequently used to transmit and receive radio signals between ground stations and the artificial probes on and around other planetary bodies in the solar system \citep{DSN_history_Mudgway:2000:}. The beams of transmitted signals are often powerful and wide enough to wash over their target and propagate out into deep space, where they could be received with a serendipitously aligned receiver. Similarly, by definition, Earth is favorably aligned to receive powerful signals transmitted from one planetary body and washing over another during a PPO event. For 2 of the 7 PPO events found in this work, the transmission source planet is within the system's habitable zone \citep{TRAPPIST-1_Gillon:2024:}. But sources of leaked transmission is not necessarily limited to planets in the habitable zone. Satellites or ground-based relays on uninhabited worlds could be setup and used to communicate for reasons not limited to our own orbiters and rovers on other planets within the solar system. If powerful transmissions are being sent between planetary bodies during these events, wide enough to not be entirely blocked by their intended target and catching the Earth within the opening angle of the propagating beam, this unintentionally ``leaked" radio emission can be detected as a technosignature \citep{siemionGHzSETISurvey2013, SETI_review_Wright:2021:}.

The purpose of this paper is to:
\begin{enumerate}
    \item Report the results of a search for narrowband radio technosignatures from the TRAPPIST-1 system.
    \item Demonstrate an application of the PPO method to observational data.
    \item Describe the application of and current state of the \texttt{NbeamAnalysis} pipeline.
\end{enumerate}

We used the newly upgraded Allen Telescope Array (ATA) at the Hat Creek Radio Observatory (HCRO), to observe the TRAPPIST-1 system for 28 hours from 0.9 to 9.3 GHz. We subsequently analyzed the data for narrowband radio technosignatures over the entire data set and then more closely examined it during predicted PPO events within the observational windows. The details of the observations are discussed in \S \ref{sec:observations_and_data}. The pipeline created to analyze the data is detailed in \S \ref{sec:data_analysis}. The results of the analysis are highlighted in \S \ref{sec:results}. And finally, \S \ref{sec:discussion} concludes with a discussion of the project and future work. 

\section{Observations and Data}
\label{sec:observations_and_data}
\medskip

The ATA is a 42-element radio interferometer located in Hat Creek, California at the Hat Creek Radio Observatory (HCRO). It is comprised of 6.1~m offset Gregorian antennas, and is currently in the process of being upgraded with new ``Antonio'' feeds: dual-polarization log-periodic feeds which are sensitive from 1--11~GHz and cryogenically-cooled to 70~K. The analog signals from the dishes are sent to the HCRO signal processing room over optical fiber, where they are mixed with four local oscillators (LOs) to produce four independently-tuned signal chains. After digitization (and fringe rotation and phase centering to the center of the primary beam), the signals are ingested by the digital signal processing backend, which provides a correlator, for imaging and beamformer calibration, and a beamformer. 

\begin{figure}
    \centering
    \begin{minipage}[hbt!]{\textwidth}
        \centering
        \includegraphics[width=\textwidth]{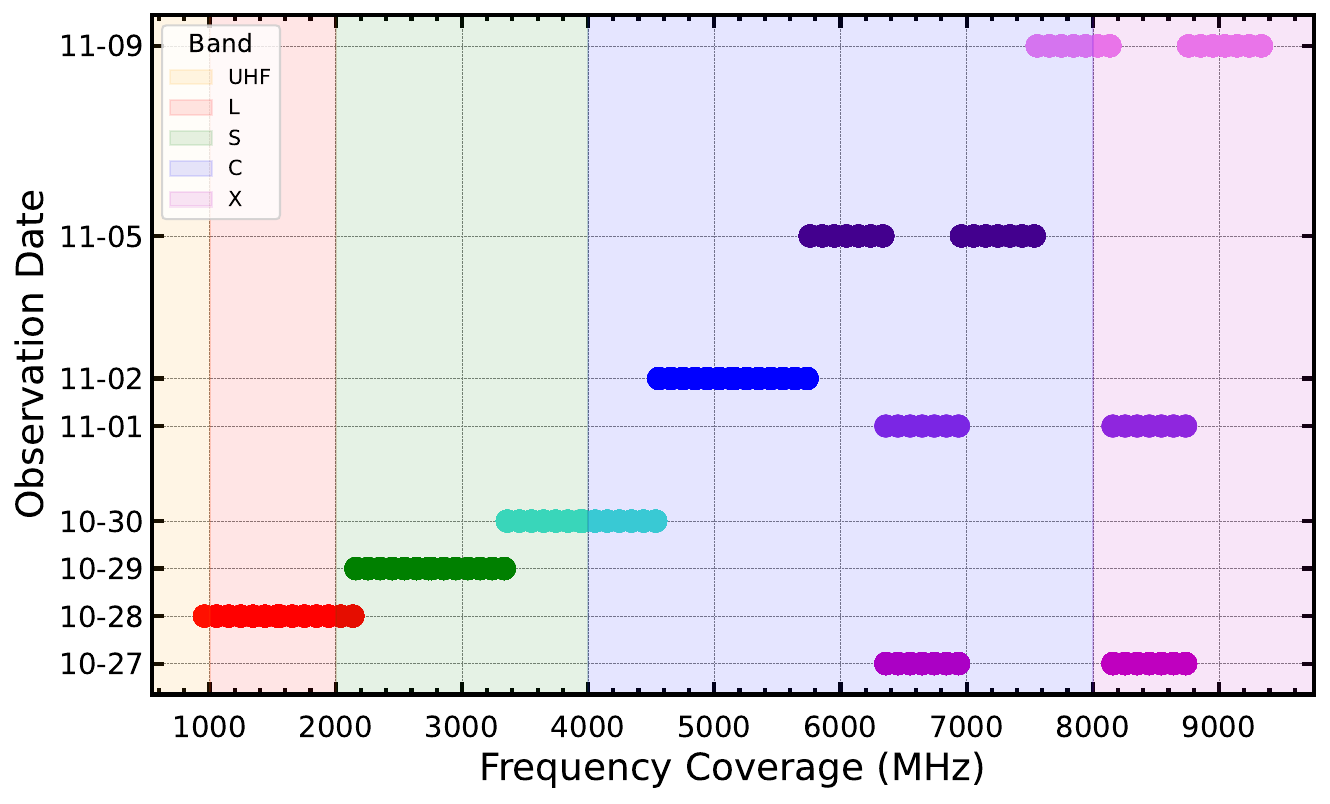}
        \caption{Frequency coverage (x-axis) by day (y-axis) of the TRAPPIST-1 system observations using the Allen Telescope Array at Hat Creek Radio Observatory in Fall 2022. The standard radio frequency bands are shaded in the background and labeled. These observations covered a continuous 0.9--9.3~GHz in radio frequency, over two weeks in October--November 2022.}
        \label{fig:obs_plot}
    \end{minipage}
\end{figure}

\begin{deluxetable}{|c|c|c|c||c|c|c|c|c|c|}
\tabletypesize{\footnotesize}
\tablecolumns{10}
    \tablehead{\multicolumn{10}{c}{2022 ATA Observations of TRAPPIST-1}} 
        \startdata
        \multirow{2}{*}{Date (UTC)} & Frequency & \multirow{2}{*}{Duration} & Total & PPO & Event & PPO & Dish & Transmitter & \multirow{2}{*}{EIRP} \\
        & Ranges (GHz) & &  Hits\textsuperscript{a} & Events & Duration & Hits & Diameter & Power Range & \\
        \hline
        \multirow{2}{*}{Oct 27, 2022} & 6.3 - 6.9 & \multirow{2}{*}{\SI{200}{\minute}} & \multirow{2}{*}{6767} & \multirow{2}{*}{de} & \multirow{2}{*}{\SI{23.0}{\minute}} & \multirow{2}{*}{38} & \multirow{2}{*}{3.4\textsuperscript{c} m} & \multirow{2}{*}{0.023 - 0.72 GW} & \multirow{2}{*}{2.17 - 69.1 TW} \\
        & \& 8.1 - 8.7 & & & & & & & & \\
        \hline
        \multirow{2}{*}{Oct 28, 2022} & \multirow{2}{*}{0.9 - 2.1} & \multirow{2}{*}{\SI{240}{\minute}} & \multirow{2}{*}{2167448} & cf, & \SI{54.7}{\minute}, & 812, & 5.89 m, & 0.15 - 2.43 GW, & \multirow{2}{*}{2.55 - 40.8 TW} \\
        & & & & cd & \SI{77.8}{\minute} & 694 &  3.4 m & 0.46 - 7.29 GW & \\
        \hline
        \multirow{2}{*}{Oct 29, 2022} & \multirow{2}{*}{2.1 - 3.3} & \multirow{2}{*}{\SI{240}{\minute}} & \multirow{2}{*}{774082} & ge, & \SI{95.0}{\minute}, & 537, & 4.81 m, & 0.15 - 3.32 GW, & \multirow{2}{*}{4.12 - 91.7 TW} \\
        & & & & df\textsuperscript{b} & - & - & - & - & \\
        \hline
        Oct 30, 2022 & 3.3 - 4.5 & \SI{240}{\minute} & 3187921 & gb & \SI{60.5}{\minute} & 358 & 7.6 m & 0.065 - 1.48 GW & 8.40 - 190 TW \\
        \hline
        \multirow{2}{*}{Nov 01, 2022} & 6.3 - 6.9 & \multirow{2}{*}{\SI{40}{\minute}} & \multirow{2}{*}{1147} & \multirow{2}{*}{-} & \multirow{2}{*}{-} & \multirow{2}{*}{-} & \multirow{2}{*}{-} & \multirow{2}{*}{-} & \multirow{2}{*}{4.28 - 137 TW} \\
        & \& 8.1 - 8.7 & & & & & & & & \\
        \hline
        Nov 02, 2022 & 4.5 - 5.7 & \SI{240}{\minute} & 27903 & hd & \SI{37.4}{\minute} & 21 & 6.8 m & 0.018 - 0.57 GW & 2.9 - 92.8 TW \\
        \hline
        \multirow{2}{*}{Nov 05, 2022} & 5.7 - 6.3 & \multirow{2}{*}{\SI{240}{\minute}} & \multirow{2}{*}{24422} & \multirow{2}{*}{bc} & \multirow{2}{*}{\SI{8.6}{\minute}} & \multirow{2}{*}{7} & \multirow{2}{*}{3.4\textsuperscript{c} m} & \multirow{2}{*}{0.090 - 2.88 GW} & \multirow{2}{*}{6.45 - 207 TW} \\
        & \& 6.9 - 7.5 & & & & & & & & \\
        \hline
        \multirow{2}{*}{Nov 09, 2022} & 7.5 - 8.1 & \multirow{2}{*}{\SI{240}{\minute}} & \multirow{2}{*}{5157} & \multirow{2}{*}{-} & \multirow{2}{*}{-} & \multirow{2}{*}{-} & \multirow{2}{*}{-} & \multirow{2}{*}{-} & \multirow{2}{*}{13.3 - 421 TW} \\
        & \& 8.7 - 9.3 & & & & & & & & \\
        \hline
        \hline
        Total: & \SI{0.9}-\SI{9.3}{\giga\hertz} & \SI{28}{\hour} & 6194847 & 7 & \SI{6.0}{\hour} & 2467 & - & - & 2.17 - 421 TW \\
        \enddata
    \vspace{0.1cm}
    \textsuperscript{a} The number of signals detected by turboSETI utilizing frequency scrunching and subsequently surviving spatial filtering across beam-formed beams.\\
    \textsuperscript{b} This potential event was not counted, though a slightly smaller dish would have created a wider beam that would have triggered an event, as explained in \S \ref{NbodyGradient}.\\
    \textsuperscript{c} The assumed dish diameter in these events is larger than the smallest dish needed in the most extreme misalignment scenario due to uncertainties described in \S \ref{NbodyGradient} and illustrated in Figure \ref{fig:PPO_sketch_2}. However, it is much smaller than the smallest dish needed in the most favorable alignment of the extreme case, and these events occur at less geometrically extreme orbital positions.
    \caption{Observations of the TRAPPIST-1 system using the Allen Telescope Array at Hat Creek Radio Observatory in Fall 2022, including: the start date of the observations in UTC; the range of frequencies covered by both tunings; the duration of the observations; the total number of signals detected by the \texttt{turboSETI} code described in Section \ref{sec:data_analysis}; the PPO events that occurred within the observation window; the PPO events respective durations; the number of candidate hits---above the nominal cutoff as described in \S \ref{Mars}---that occurred during the predicted event windows; the dish diameter assumed for triggering an event; and the minimum transmitter power these observations would be sensitive to. PPO events are listed as the occulted body followed by the occulting body, and were determined using modeling from \texttt{NbodyGradient} from \cite{NbodyGradient_Agol:2021:} as described in \S \ref{NbodyGradient}. The minimum transmitter power was calculated for each assumed dish diameter at the observed frequency, and the range shows the degredation of sensitivity at higher drift rates, even with frequency binning to recover power loss as explained in \S \ref{sec:signal_detection} \citep{fscrunch_power_loss_Sheikh:2023:}. The EIRP was calculated as discussed in section \S \ref{sec:results} using the maximum SEFD, with the range indicating the degradation in sensitivity at increasing drift rates.} 
    \label{tab:observations}
\end{deluxetable}

For this project, signals from 20 dishes were digitized by RFSoC boards using two tunings, leading to the ability to record two 672 MHz tunings simultaneously. The beamformer was configured to create two spatially separated beams within the primary field of view, the on-beam placed at phase-center of the array and centered on the target, and the off-beam fixed at a constant 9 arcminutes away, producing high-frequency resolution filterbank data files for each beam \citep{SIGPROC_Lorimer:2011:}. The FWHM of the tied beam varies from $\sim3$ arcminutes at the bottom part of the observed frequency band, to $\sim0.5$ arcminutes at the top part. This corresponds to a separation of $\sim 3-18\times$\,FWHM between the on and off-beams, and an attenuation of at least 10\,dB between them according to beam-pattern measurements.
The entire frequency range of the data was recorded in this way over 8 compute nodes with a coarse channel size of 0.5 MHz, a narrow channel resolution of 1 Hz and binned to 16 s over every integration, yielding a total data volume of 65 TB. The ATA RFSoC boards perform the coarse channelization using a polyphase filterbank before streaming the channlized voltages to the acquisition servers. Fine channelization is achieved by running a GPU-accelerated real-time $\sim$500k-point FFT on the data acquisition servers.

The ATA was used to search the TRAPPIST-1 system for narrowband radio emission over the dates and tunings shown in Table \ref{tab:observations} and Figure \ref{fig:obs_plot}. Each observation was conducted in 10 minute integrations for a total of 240 minutes per frequency band, one of which had to be broken up between 2 different observation days to be completed. These are treated separately in the analysis to account for temporal differences in the RFI environment.
\section{Data Analysis}
\label{sec:data_analysis}


\subsection{Signal Detection with turboSETI}\label{sec:signal_detection}

\begin{figure}
    \centering
    \begin{minipage}[hbt!]{\textwidth}
        \centering
        \includegraphics[width=\textwidth]{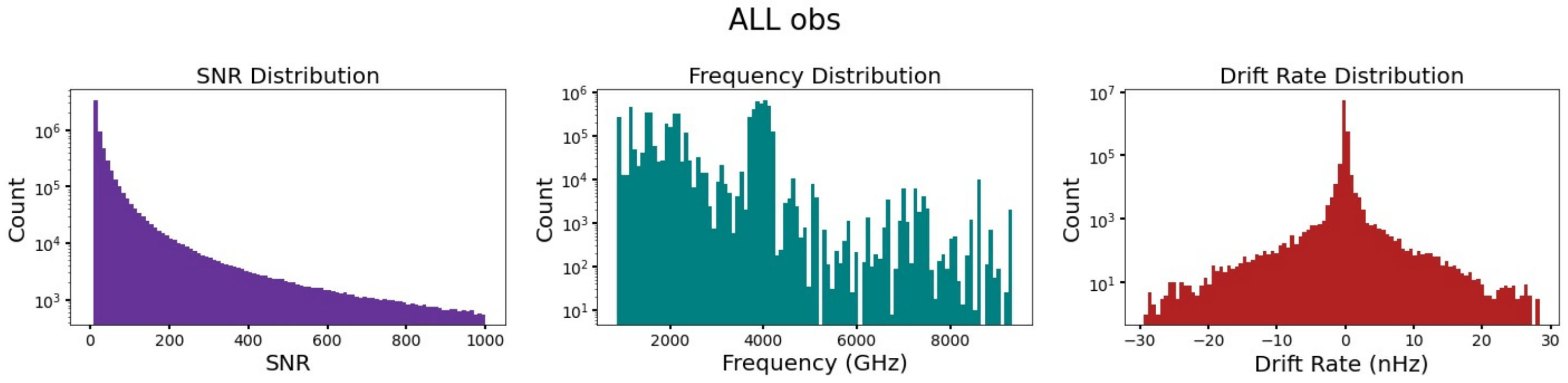}
        \caption{Distributions of the relevant signal parameters from all 8 observations combined. The SNR is truncated at 1000 for readability, but the data extends out several orders of magnitude, up to a maximum of $\sim1.15\times10^7$ for a few ($\sim$1\% of the data) powerful nearby signals. Most signals were detected at lower frequencies where the major bands are known to be more crowded with RFI. The drift rate distribution extends beyond the target maximum of 15 nHz due to bin doubling during the ``frequency scrunching'' process, but is still strongly clustered around 0 nHz due to an abundance of terrestrial RFI.}
        \label{fig:stats_all}
    \end{minipage}
\end{figure}

The tree de-Doppler algorithm from \texttt{turboSETI}, an open-source software package developed by Breakthrough Listen \citep{Enriquez_2017ApJ...849..104E}, is the primary publicly available tool currently being used to search for drifting narrowband signals in high-frequency resolution filterbank data. We applied the algorithm to the TRAPPIST-1 data to identify signals or hits in both beams above a signal to noise ratio (SNR) of 10. Figure \ref{fig:stats_all} shows diagnostic histograms for all of the signals detected over all observations, with the maximum SNR truncated to 1000 for readability (1\% of the $\sim$6 million signals were above SNR of 1000, with the largest SNR value above 11.5 million). The observations covered different frequency ranges, leading to widely-varying RFI environments and, thus, numbers of hits. Unsurprisingly, the lower frequency bands, L and S, tended to have more hits than the higher frequencies, C and X.

Without knowing the transmitted frequency, a Doppler shift alone is degenerate between a transmitter frequency offset and a non-zero line-of-sight velocity between the source and the observer. However, the ``drift rate'' of a signal measures the rate at which a transmitted signal shifts through frequency space as it is observed on Earth due to the changing relative motion between the source and the observer. The drift rate is therefore more useful in identifying signals from distant sources with an expected acceleration relative to us as observers. The de-Doppler algorithm used by \texttt{turboSETI} calculates this drift rate in each signal it identifies. One limitation of this de-Doppler technique is its sensitivity loss at higher drift rates than the one-to-one pixel rate of the data due to power smearing over multiple pixels at higher drift rates \citep{Enriquez_2017ApJ...849..104E}. We configured the ``beamformer mode-h'' data produced by the Breakthrough Listen Accelerated DSP Engine (BLADE) back-end to have a frequency resolution of 1 Hz and a time resolution binned to 16 seconds. Thus, any signals with drift rates higher than 1/16 Hz s$^{-1}$ will have their power smeared over multiple pixels and the sensitivity of the tree summation in the deDoppler algorithm will be significantly reduced. The loss in sensitivity at these higher drift rates scales roughly as $1 - \alpha/x$, where $\alpha$ is the pixel ratio, 1/16 Hz s$^{-1}$ in this case, and $x$ is the actual drift rate of the signal \citep{Margot_explains_BL_sensitivity_loss:2021:}.

Radio SETI searches must generally remain agnostic about the relative motion of the transmitter source, especially for systems without known planets. It has been shown that planetary systems with high accelerations, e.g., those with tightly orbiting planets or highly eccentric orbits, could produce narrowband signals with large drift rates due to their relatively high orbital motion, thus complicating the search for such signals \citep{MDR_Sofia_2019ApJ...884...14S,Li:2023:}. The recommendation from \citet{MDR_Sofia_2019ApJ...884...14S} suggested that drift rates up to 200 nHz (1 kHz s$^{-1}$ at 5 GHz) should be considered in searches targeting stars with little prior knowledge on the orbital characteristics of planets within a given system. (The unit of nHz was there developed and has since gained traction in the field as a frequency-independent unit of drift rate and a measurement of the acceleration over the speed of light, a/c. The more traditional drift rate units of Hz s$^{-1}$ are recovered by multiplying a/c, in nHz, by the observed frequency in GHz.)

However, it is assumed that the TRAPPIST-1 planets are tidally locked due to their proximity to their host star and will have a negligible rotational contribution to the drift rate of a transmitter on their surfaces. Additionally, their orbital parameters are well constrained, making it possible to calculate the drift rate contributions from their orbital motion. Satellite transmitters in circular orbit around each planet could produce much higher drift rates, up to an additional $\sim$45 nHz on top of the contribution from the planet's orbit around the star. However, we have chosen to limit our scope to analogues of our deep space communications, the strongest of which are surface transmitters to deep space probes. Although our space probes and satellites have transmitters that return data to Earth, these are typically significantly weaker in power, making it much more practical to search first for more powerful surface transmitter signals. Rather than calculating the drift rate from a particular body based on the precise orbital configuration during the time of each observation, we determined a maximum drift rate of 15 nHz for a surface transmitter in the TRAPPIST-1 system using the methodology from \citep{Li:2023:} for the innermost planet --- this was used as a conservative value to ensure that our search is sensitive to accelerations from transmitters on any of the planetary surfaces in the system. A more rigorous calculation of the anticipated drift rate range during a PPO event and subsequent data analysis would be warranted in the event of an interesting signal candidate, but this extra consideration ultimately proved unnecessary.

Even at the lower end of the frequency range of $\sim$1 GHz for these observations, the resulting maximum drift rate of $\sim$15 Hz s$^{-1}$ is far above the 1/16 Hz s$^{-1}$ pixel ratio where power loss could become a concern. In an attempt to recover some of the power loss during signal detection for these higher drift rates, we used a technique called ``frequency scrunching'' or ``fscrunching.'' Similar to the methods used by \cite{siemionGHzSETISurvey2013} and \cite{fscrunch_power_loss_Sheikh:2023:}, fscrunching bins power over frequency channels to increase the one-to-one pixel ratio of the data, as described below. 

We first ran the de-Doppler algorithm over an absolute value drift rate range of 0 to 1/16 Hz s$^{-1}$. Although we typically only refer to positive drift rate ranges, the algorithm mirrors these values in the negative direction and searches the equivalent negative drift rate range as well, so it should be understood that the negative value range is included in the search. Then fscrunch was applied to bin adjacent frequency channels together, doubling the pixel ratio, and the deDoppler search was run again from 1/16 to 1/8 Hz s$^{-1}$. This doubling of frequency bins was continued until the maximum drift rate of 15 nHz was covered for all frequencies. Up to 12 iterations of this technique were needed to include the maximum drift rate of 139.5 Hz s$^{-1}$ at the highest frequency of 9.3 GHz, extending the search out to an actual maximum of 256 Hz s$^{-1}$ or 27.5 nHz. The minimum number of iterations was 8 at the lowest frequency.
The distribution of drift rates can be seen in Figure \ref{fig:stats_all}, although the vast majority of signals are found closer to 0 Hz s$^{-1}$, as expected for a local RFI-dominated environment.


\subsection{NbeamAnalysis Pipeline}

We developed the \texttt{NbeamAnalysis}\footnote{\url{https://github.com/SETIatHCRO/ATA-Utils/tree/master/NbeamAnalysis}} pipeline as an open-source filtering tool for signals identified in beamformed filterbank data products at the ATA. Similar in purpose to the post-signal detection filtering functions of \texttt{turboSETI}, the \texttt{NbeamAnalysis} pipeline produces a filtered list of signals with their relevant parameters along with overall diagnostic plots for quick review of the results. The resulting list of filtered signals can then be fed into a versatile plotting tool to plot the dynamic spectra of individual signals for visual comparison across beams. The pipeline is designed to handle any number of beam-formed data products, though the data volume for N$>$2 may be prohibitive. 


\subsubsection{Spatial Filtering}\label{sf}

The search strategy identified $>$24 million signals across all observations. However, the majority of these hits were found with similar power at the same frequency (within 2 Hz) over the same frequency range in both beams. This indicates that they are likely local RFI, washing over the entire field of view, and not spatially distinct within the individual beams. Due to the sparse array configuration of the 20-element ATA, the synthesized beam possesses high sidelobe levels. 
Therefore, even a true sky-localized signal could appear in the off-target beam with only moderate power attenuation. From both analytical estimates and empirical tests on sky-localized X-band downlinks from orbiters around Mars (see Section \ref{Mars}), the minimum attenuation factor between the on- and off-beam was determined to be roughly 4. This attenuation requirement provides an additional RFI filter. By cross-referencing each hit in each beam, we identify hits having the same frequency as well as attenuation factors less than this factor of 4 as RFI. These hits were removed from further consideration and we refer to this rejection method as ``spatial filtering.'' Execution of the spatial filtering process reduced the total number of hits to $\sim$6 million for further processing.


\subsubsection{DOT}\label{DOT}

We developed and incorporated the DOT algorithm within the \texttt{NbeamAnalysis} pipeline as a novel filtering technique to identify candidate signals after the optional spatial filtering process. The algorithm is used to score the correlation of identified signals between multiple beam-formed beams by comparing the power data matrices in each beam over the frequency range for that signal. Under the assumption that signals from a distant source should be isolated within the target beam and attenuated in any spatially separated beams, the expectation is that hits with lower DOT scores are more likely to be localized within a single beam and thus contain interesting signals for further follow up. 

From a given filterbank file, data may be extracted as a sliced subset of the larger filterbank file over a defined frequency range, and represented in a power matrix, where the rows are time bins, $\sim$16 s, the columns are frequency bins, $\sim$1 Hz, and the values measure the power in each cell. The score between any two data slices, $x$, is calculated with a normalized dot product of the data slice matrices,

\vspace{-0.3cm}
\begin{equation}
    x = \frac{\sum_{i,j}^{N,M} A_{i,j} B_{i,j}}{\sqrt{\sum_{i,j}^{N,M} A^2_{i,j} B^2_{i,j}}},
\end{equation}

where $A$ and $B$ are the power data matrices of the same shape, $N \times M$. This function assumes that the median of the background noise is zero. To enforce this, the same assumption used in \texttt{turboSETI} to determine the median noise level is utilized. In a given data slice, data below the 5th and above the 95th percentile of power are removed and the median is taken from the remaining data. 
It is additionally assumed that the noise profile is roughly the same in both beams. In this way, a signal that appears in one beam and not the other should be scored close to zero, while most RFI appearing with the same power profile in both beams should be scored close to one. 

\subsubsection{Injection Recovery}
\label{setigen}

To test the effectiveness of scoring signals with the DOT algorithm, we used \texttt{setigen} to inject artificial signals into the data \citep{setigen_Brzycki:2022:}. As a preliminary test, three signals were injected; The first was a strong signal injected in a region of pure noise; The second was injected on top of RFI with a signal strength much greater than the RFI signal detected by \texttt{turboSETI}; The third was injected on top of RFI with a signal strength comparable to the RFI signal strength. The pipeline was then run on the data with these injected signals.

\begin{figure}
    \centering
    \begin{minipage}[hb!]{\textwidth}
        \centering
        \includegraphics[width=\textwidth]{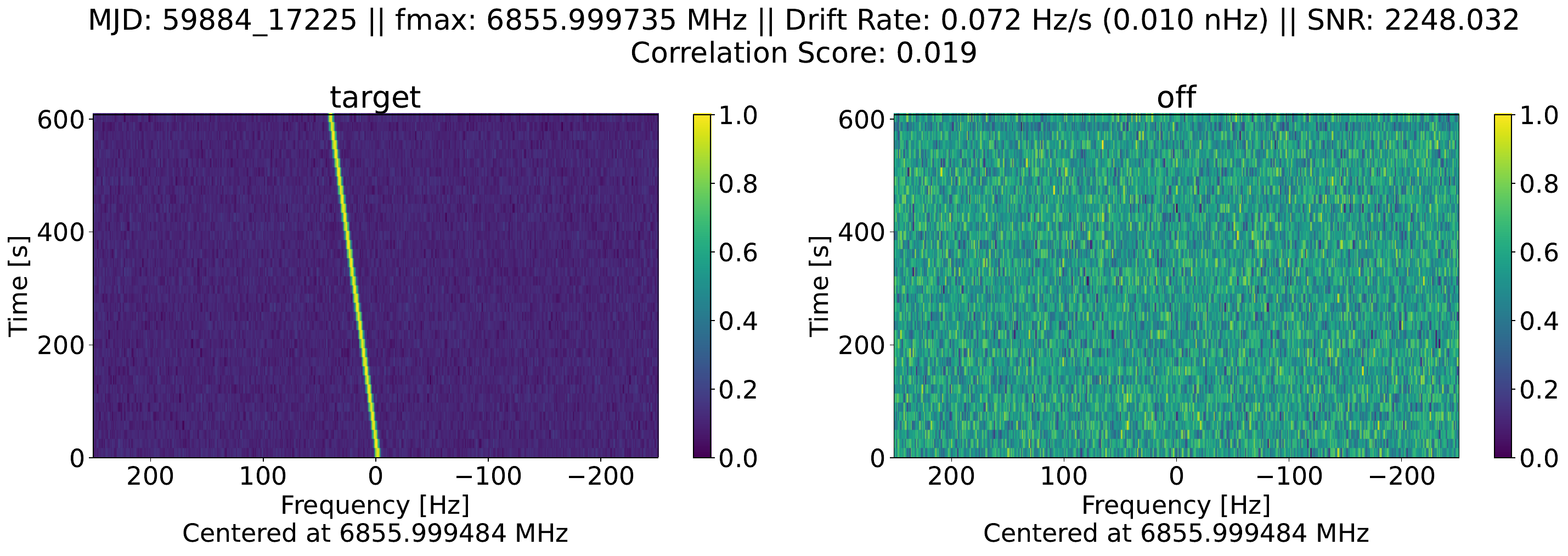}
        \caption{An injection-recovery test using \texttt{setigen} used to score an isolated signal with the DOT algorithm. The isolated linearly-drifting signal seen in the target panel was injected in a region of frequency space known to contain only noise. The signal was given a width of 2 Hz, a drift rate of 0.07 Hz s$^{-1}$, and an SNR of 42,000. \texttt{turboSETI} calculates SNR independently from the SNR input into \texttt{setigen}, but nevertheless indicates a relatively strong signal as intended. The low correlation score of 0.019 indicates that the signal detected in the target panel is distinct from the pure noise in the off panel.}
        \label{fig:injection1}
    \end{minipage}
\end{figure}

\begin{figure}
    \centering
    \begin{minipage}[hb!]{\textwidth}
        \centering
        \includegraphics[width=\textwidth]{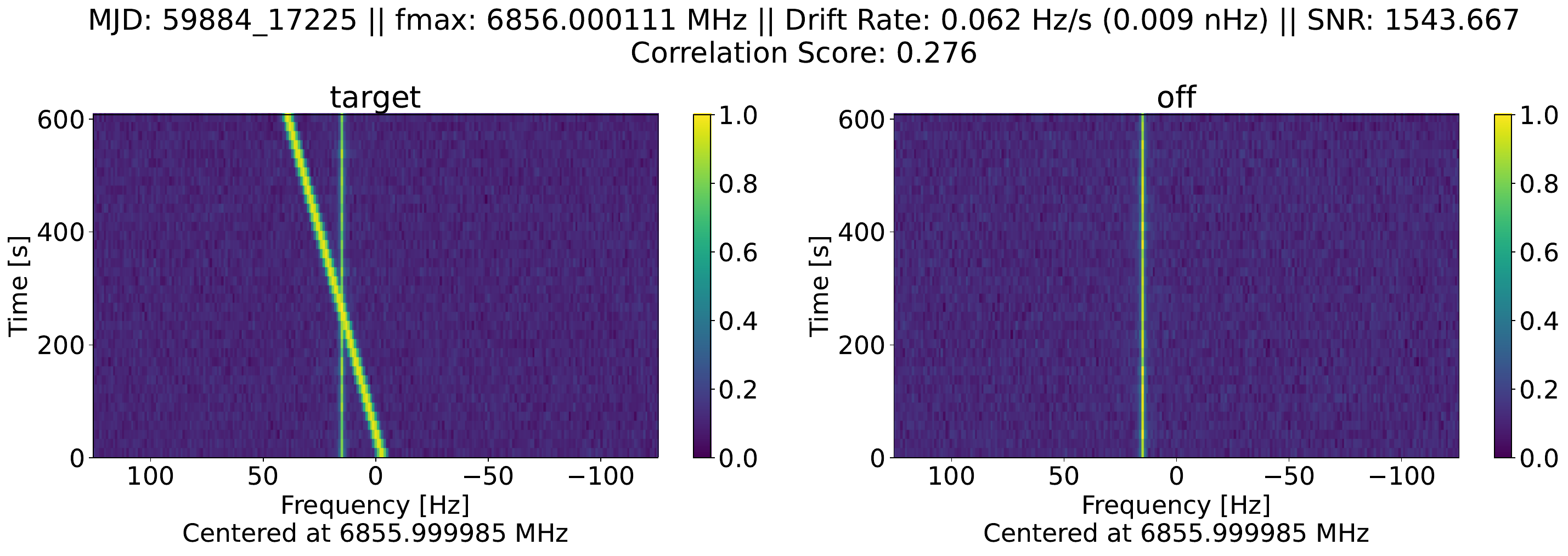}
        \caption{The second injection recovery test using \texttt{setigen} to score a strong signal on top of weaker RFI. The added signal seen only in the target panel was injected on top of previously identified RFI, originally calculated by \texttt{turboSETI} as 42.25. The signal was given a width of 2 Hz, a drift rate of 0.07 Hz s$^{-1}$, and an SNR of 1000 times that of the RFI. \texttt{turboSETI} calculates SNR independently from SNR input into \texttt{setigen}, but nevertheless indicates a relatively strong signal as intended. The RFI offset from 0 in these frames shows why this was not removed by spatial filtering. The somewhat low correlation score of 0.276 could indicate an interesting signal for follow-up, depending on the user-defined cutoff value.}
        \label{fig:injection2}
    \end{minipage}
\end{figure}

The injection test indicates that a narrowband signal appearing in only one beam should produce scores very near 0, unless overlapping with much stronger RFI. Figure \ref{fig:injection1} shows the resulting waterfall plot of the two beams where a signal has been injected in a region devoid of obvious RFI in the target beam. The resulting correlation (DOT) score very near 0 demonstrates that a strong signal localized in one beam of a beamformed data product is easily identifiable. 

Typically, narrowband RFI shows up in both beams and produces correlation scores significantly closer to 1 than 0, even for signals with low SNR. Figure \ref{fig:injection2} shows that even for a powerful signal injected on top of relatively weaker RFI, the pipeline can recover a relatively low score that may indicate to the user that further investigation is warranted.

\begin{figure}
    \centering
    \begin{minipage}[hbt!]{\textwidth}
        \centering
        \includegraphics[width=\textwidth]{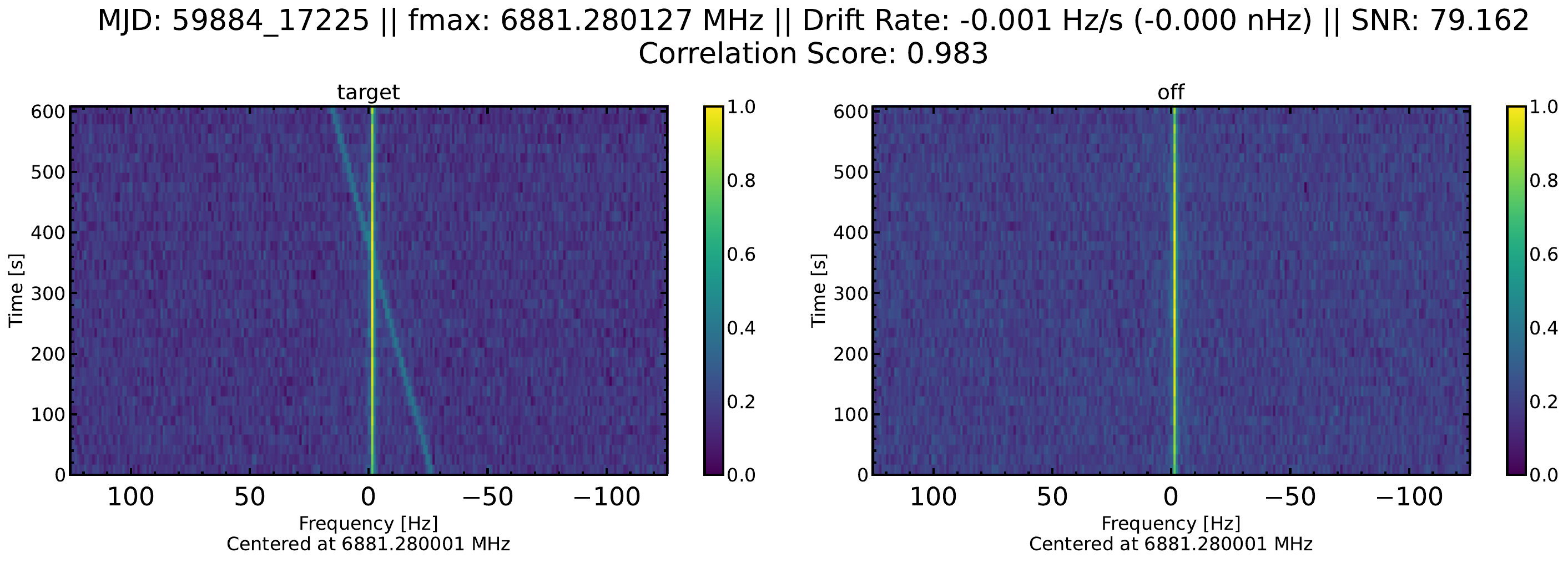}
        \caption{The third injection recovery test using \texttt{setigen} to score a weak signal on top of stronger RFI. The added signal seen only in the target panel was injected on top of previously identified RFI, originally calculated by \texttt{turboSETI} as 78.88. The signal was given a width of 2 Hz, a drift rate of 0.07 Hz s$^{-1}$, and an SNR equal to that of the RFI. \texttt{turboSETI} calculates SNR independently from the SNR input into \texttt{setigen}, but the reported SNR is relatively close to the original RFI. The RFI centered on 0 in these frames shows why this was removed by spatial filtering. The very high correlation score of 0.983 is unlikely to indicate an interesting signal for follow-up.}
        \label{fig:injection3}
    \end{minipage}
\end{figure} 

However, in the case of a signal comparable or weaker in power than the RFI it overlaps, as in Figure \ref{fig:injection3}, the stronger RFI signal is keyed in on by \texttt{turboSETI} and the hits in both beams over this region are registered at the same frequencies. Thus, if spatial filtering is applied in the pipeline, this signal could be filtered out. Even without spatial filtering, the power from the RFI dominates the data slice and the resulting score is so close to 1 that the score alone is unlikely to flag this for further investigation. The potential for true signals to be lurking in frequency space rejected by RFI is a pervasive problem in these kinds of searches. Recent attempts at employing machine-learning techniques have seen some success and may prove to be an effective tool in overcoming this problem \citep{Ma:2023:, ML_Ma:2024:}.


\begin{figure}
    \centering
    \begin{minipage}[hb!]{\textwidth}
        \centering
        \includegraphics[width=\textwidth]{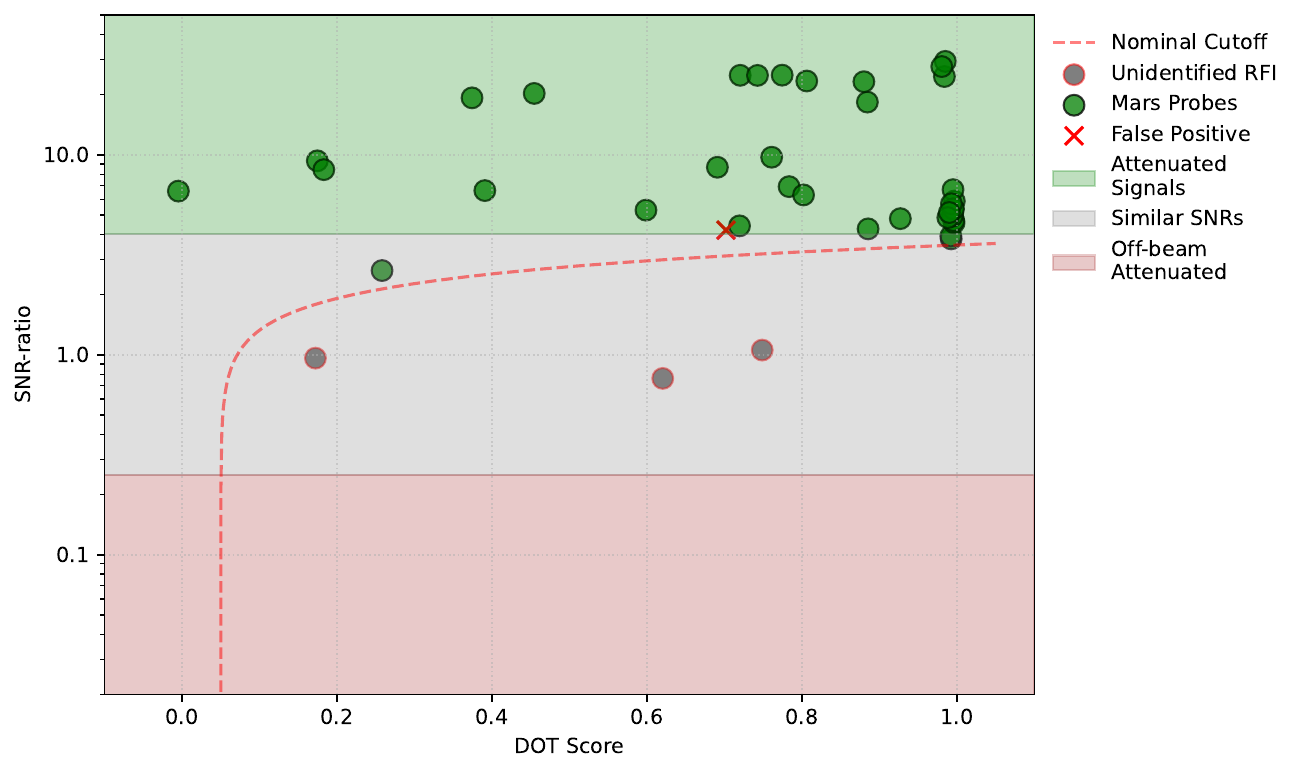}
        \caption{Results of Mars test observations showing recovery of 37 Mars orbiter downlinks during the time of observation. 3 signals were classified as RFI below the nominal cutoff, and 1 false positive showed an attenuated signal at a known transmitting frequency for STEREO-A, which was not in either beam at the time. Signals that are powerful enough to show up in both beams, overwhelming the DOT scoring algorithm and trending toward 1.0, are still attenuated in the off-beam and identified as potentially interesting for follow-up. Candidate signals localized to a distant source may show up only weakly in the target beam at the limits of the telescope's sensitivity, and are expected to appear in the top left of this kind of plot.}
        \label{fig:mars}
    \end{minipage}
\end{figure}

\subsubsection{Mars Probes} \label{Mars}

\begin{figure}
    \centering
    \begin{minipage}[hbt!]{\textwidth}
        \centering
        \includegraphics[width=\textwidth]{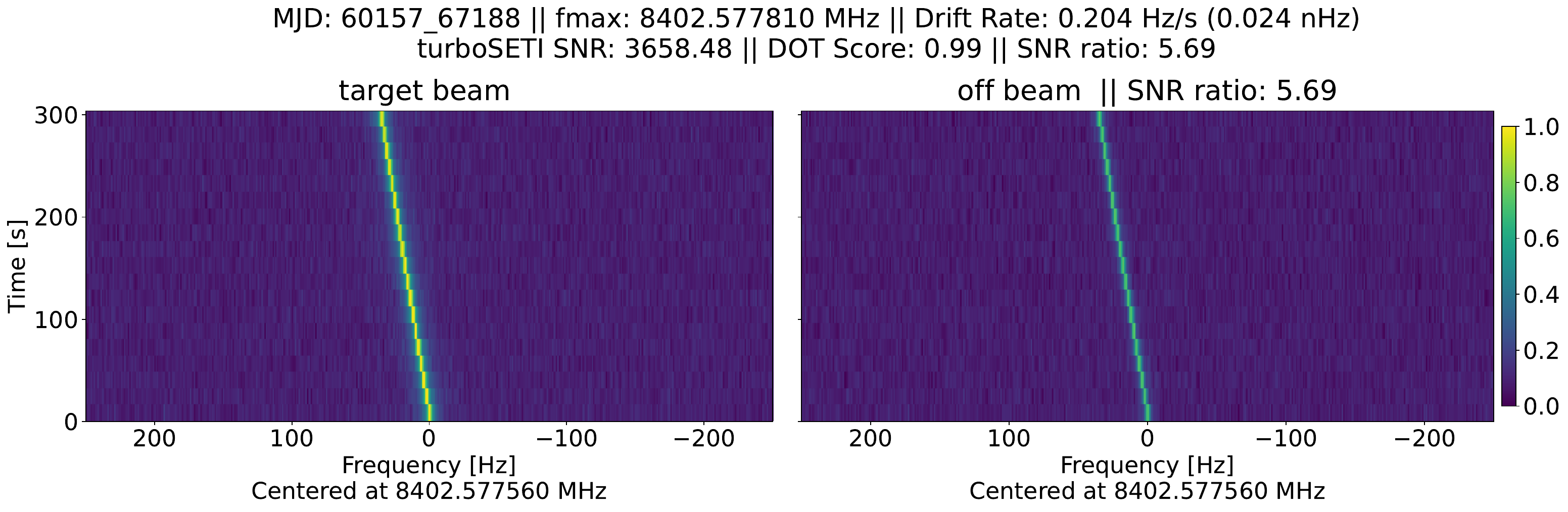} \\
        \vspace{0.5cm}
        \includegraphics[width=\textwidth]{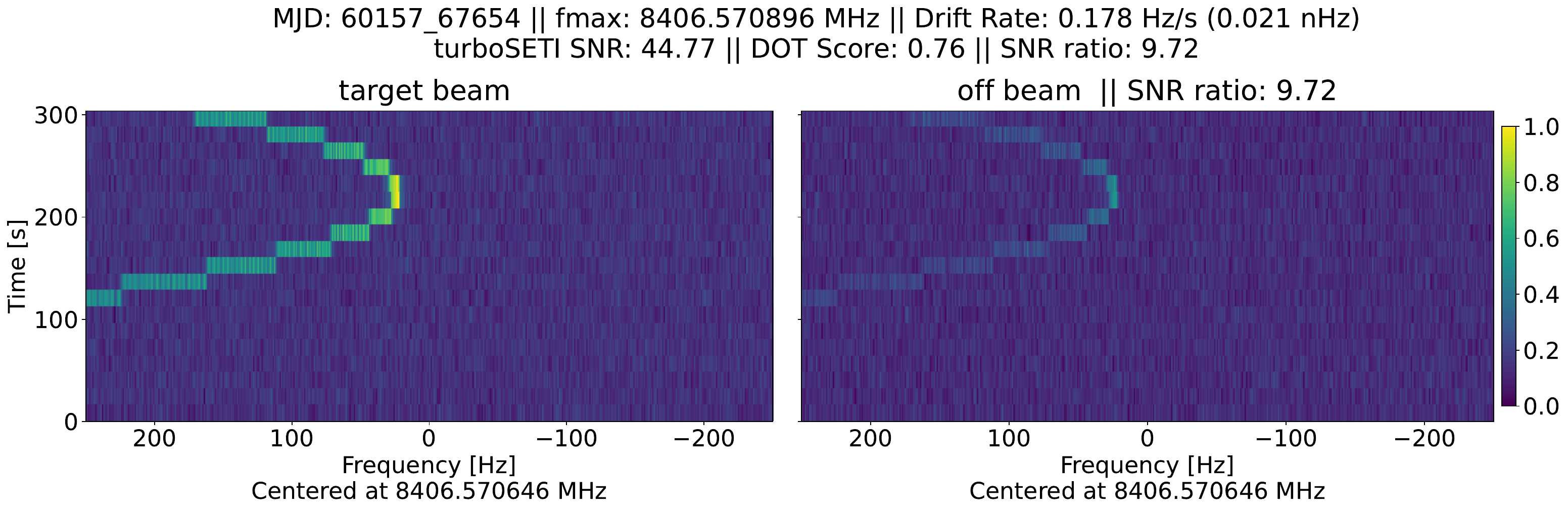} \\
        \vspace{0.5cm}
        \includegraphics[width=\textwidth]{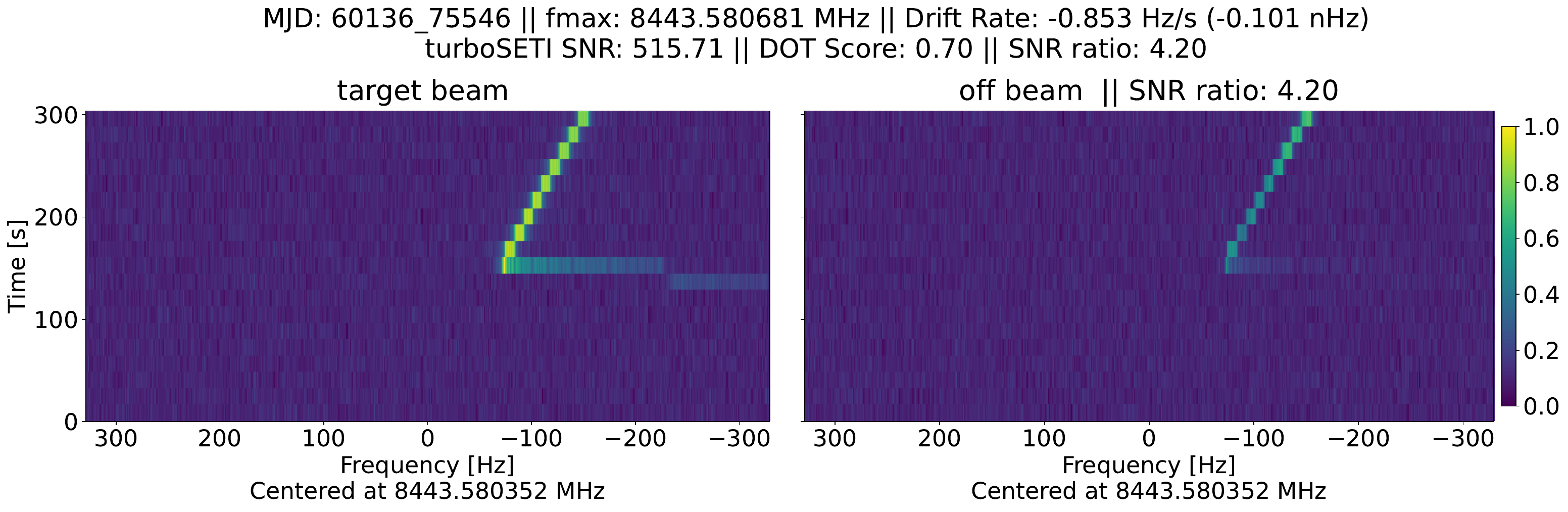}
        \caption{Three representative signals in the on- and off-beams resulting from test observations of Martian probes that were subsequently filtered through the \texttt{NbeamAnalysis} pipeline. The detected frequency of these signals were cross-referenced with known transmission frequencies for various Martian probes to identify their source. The top panel shows a signal from the Emirates Hope orbiter. The middle panel shows a signal from the Mars Odyssey orbiter. The bottom panel shows the ``False Positive" signal noted in figure \ref{fig:mars} from STEREO-A, a heliocentric satellite. The colorbar indicates the amount of power in each beam, normalized to the peak of the target beam signal and displayed in a logarithmic scale.}
        \label{fig:mars_comp}
    \end{minipage}
\end{figure} 

As an additional, practical test of the \texttt{NbeamAnalysis} pipeline, in mid-July and early August 2023, a series of high-frequency resolution observations targeting Mars were performed at the ATA utilizing the same beamforming process. Several deep space probes expected to be transmitting in the observed frequency range were catalogued for reference. The data collected during these observations were then processed with the \texttt{NbeamAnalysis} pipeline, and this test of the pipeline revealed that signals from deep space detected in the target beam were also being picked up in the off-target beam despite significant spatial separation. Through investigation of these data, we found that the application of the beamforming software was capturing power from the target beam in the off-target beam, with an attenuation factor of at least 4 between the beams. Therefore, sufficiently strong signals localized to a distant source within the target beam will not disappear in the off-target beam. 

Figure \ref{fig:mars} shows the results of including SNR-ratio (detailed in \S \ref{SNRr}) as a discriminating parameter to measure attenuation between beams. In order to isolate the bulk of expected RFI within the 2D parameter space of DOT score and SNR-ratio, while accounting for an indeterminate amount of uncertainty, we designed a nominal cutoff as a power law relation based on an empirical description of the data. We define it as: 

\begin{equation}
    y = 0.9 A (x-0.05)^{1/3},
\end{equation}

where $A$ is the attenuation value (assumed to be roughly 4 for the ATA's current configuration), $x$ is the DOT score and $y$ is the SNR-ratio. 

Discriminating the signals in this way allowed us to reject all the RFI and recover all of the true signals, including an additional ``False Positive" that turned out to be the STEREO-A spacecraft in heliocentric orbit and not in the beams at the time of observation, likely being picked up through a side-lobe. For comparison with the synthetic signals in X-band, figure \ref{fig:mars_comp} shows a few representative spectra of signals that were recovered in this test, including the additional false positive from STEREO-A.

The variation in attenuation factor for the real signals is mainly due to differences in on-sky target positions between the observations. During each observation, the off-beam is fixed at a constant relative 9 arcminute separation from the target-beam, but these results comprise several observations where the target is located at different positions on the sky, leading to variations in the observed beam pattern. Discussion of ongoing work to account for this variability in future observations can be found in section \S \ref{sec:discussion}.

\subsubsection{SNR-ratio}\label{SNRr}

Calculation of the SNR-ratio between beams for a given signal was added to the pipeline as a second major filtering parameter based on tests explained in \S \ref{Mars}. For a given signal, there is an expected attenuation value of $\sim$4 between the target and off-target beams from the beamforming process employed to create these data. A sufficiently powerful signal coming from a distant target within the target beam should show up at least 4 times weaker in the off-target beam. Thus, by calculating the SNR within every data slice that contains a hit as well as the same frequency-sliced data in the off-target beam, the SNR-ratio can be calculated and compared with the expected attenuation factor. 

To calculate the SNR of a given data slice, we measured the standard deviation of the middle 90 percent of the pixels within the data as described in \S \ref{DOT}. We identified signal in each data slice as any power greater than 10$\times$ the standard deviation of the noise. The SNR for data slices not meeting this criteria was set to 1. In data slices where signal was identified, we then calculated the signal strength to be the median of the $N$ highest power elements, where $N$ is the number of time bins, which is 38 in the 10-minute integrations of the TRAPPIST-1 data. We determined the SNR by dividing the signal by the standard deviation of the noise, and then finally, the SNR for the on- and off-beams was divided to produce the SNR-ratio for each hit.

\begin{figure}
    \centering
    \begin{minipage}[hbt!]{\textwidth}
        \centering
        \includegraphics[width=0.6\textwidth]{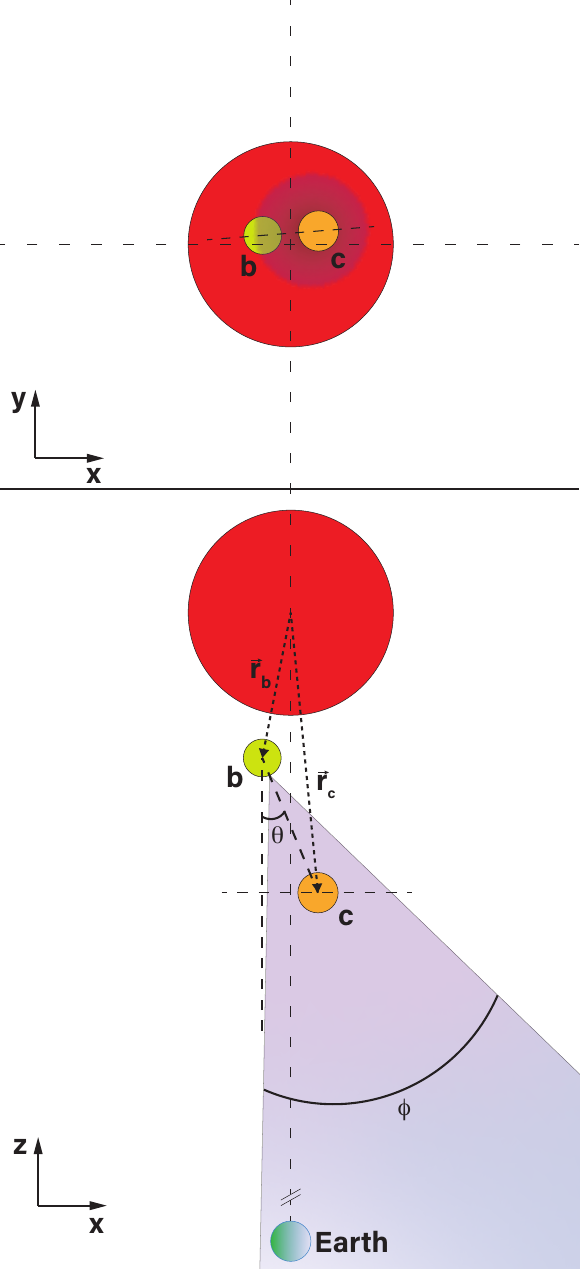}
        \caption{A cartoon illustrating that PPO events, in this work, can occur in scenarios broader than strict occultation. Specifically, PPO events occur where the opening angle of the transmitted beam, $\phi/2$, is greater than the angle $\theta$ between the line of the planets and the $-\hat{z}$ direction, toward Earth. The top panel shows the projected xy-plane view. The bottom panel shows a bird's-eye, xz-plane view of the same scenario where a transmitter on the surface of planet b is sending a radio signal toward planet c. This sketch is not to scale, but it provides a visual reference for the geometry considered in calculating a PPO event.}\label{fig:PPO_sketch_1}
    \end{minipage}
\end{figure}


\subsection{PPOs}\label{NbodyGradient}




Thorough analysis to determine the orbital parameters of the TRAPPIST-1 planets by \cite{agolRefiningTransittimingPhotometric2021} suggests that the planets in this system are very co-planar and nearly edge-on, with some small degree of offset and uncertainty. But even a small deviation in the relative inclinations of the planets can turn a syzygy into a near miss. 

\begin{figure}
    \centering
    \begin{minipage}[hbt!]{\textwidth}
        \centering
        \includegraphics[width=\textwidth]{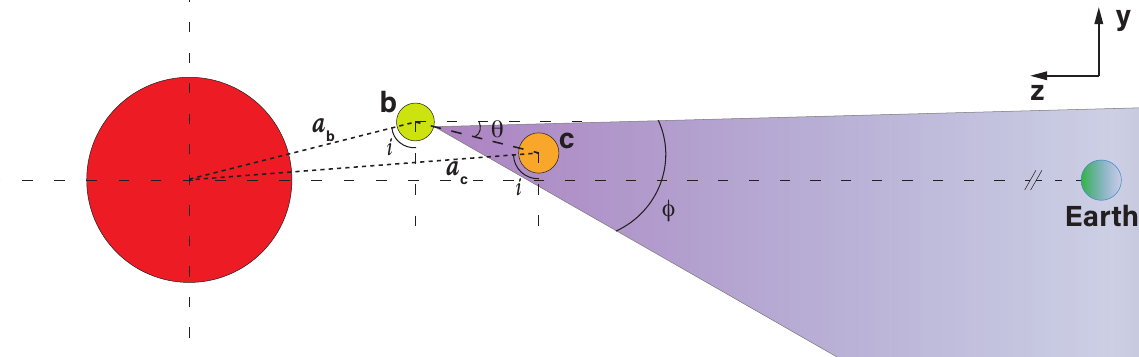}
        \caption{Sketch of the geometry of a PPO event where the effects on our observations due to uncertainties in the semimajor axis and inclination would be maximized. When the opening angle of the transmitted beam, $\phi/2$, is greater than the angle $\theta$ between the line of the planets and the $-\hat{z}$ direction, spillover emission from a PPO event could be detected.}\label{fig:PPO_sketch_2}
    \end{minipage}
\end{figure}

However, for the purposes of searching for radio spillover, we use a broader definition of PPOs that do not require syzygy. Here, a ``PPO event'' defines a window of time wherein Earth could receive radiation from a beamed radio signal from the further planet towards the nearer one. The range of valid geometries is based on the width of the opening angle, which itself depends on the frequency and dish diameter. Figures \ref{fig:PPO_sketch_1} \& \ref{fig:PPO_sketch_2} provide cartoon illustrations of favorable geometries from different points of view. 

The three points of the transmitting planet, receiving planet, and the Earth constitute a plane. In this plane, the radio transmission is beamed along a line between the two planets in an expanding cone with some opening angle, $\phi = 1.15 \lambda/D$,
where $\lambda$ is the wavelength of transmission and $D$ is the diameter of the transmitter dish. In this way, $\phi$ is defined as the FWHM of the beam. For the sake of simplicity and because we are only defining windows in which to focus our search based on notional dish sizes, we assume most of the power is contained within this angle and drops off steeply enough to be approximated as a top-hat function with the FWHM as its width. When half of that opening angle, $\phi$, is larger than the angle between the central beam line and Earth, $\theta$, there is a PPO event. In this way, syzygy is not required for a PPO event with a sufficiently large transmitted beam. 

The DSN typically uses a 34 m dish to transmit to probes on Mars. The separation of planets in the TRAPPIST-1 system is an order of magnitude smaller than the distance between Earth and Mars, meaning that dish diameters can be an order of magnitude smaller than those used by the DSN to achieve the same gain and signal strength at a given power.  
We therefore started with an assumed dish diameter of 3.4 m for communication between neighboring planets and added an additional scaling factor. The DSN uses larger dishes for communication with our deep space probes further away, so we included a scaling factor to account for the possibility of larger dishes being used to communicate to planets further apart. This scaling factor was chosen as the square root of the difference between the numbered order of each planet. This choice was made to yield reasonably sized dishes and resulting beams that essentially vary with target distance and provide a reasonable constraint on a detectable transmitting beam width. The scaled dish diameter and the maximum frequency of each observation were used to calculate the opening angle of a transmitted beam, which determined whether that triggered a PPO event. The duration of each event was calculated by totaling the number of steps in the simulation, with a resolution of 2.9 minutes, where $\phi/2 > \theta$ for the relative positions of that pair of planets with respect to Earth.

\begin{figure}
    \centering
    \begin{minipage}[hbt!]{\textwidth}
        \centering
        \includegraphics[width=\textwidth]{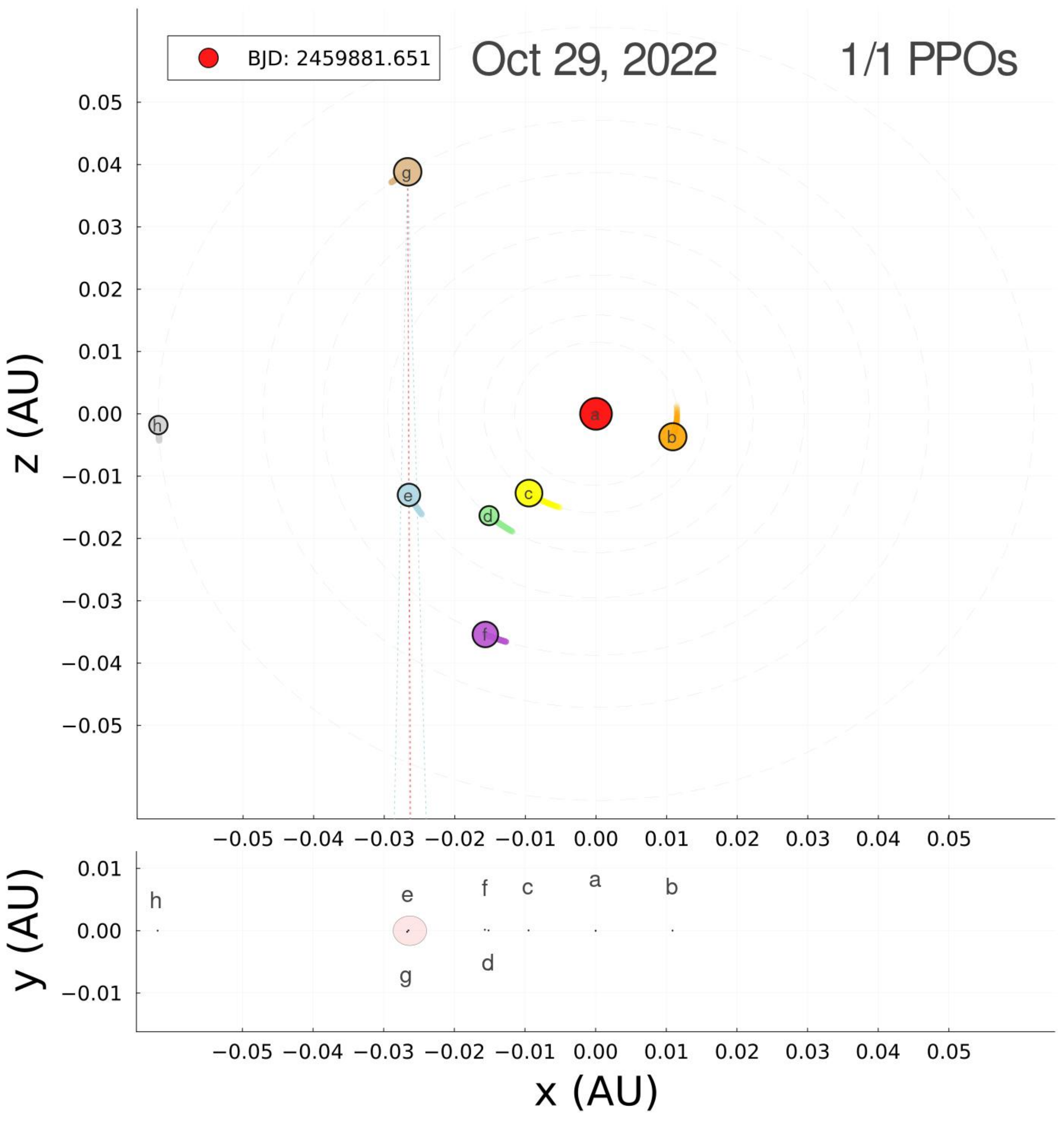}
        \caption{Simulated potential PPO events during our observations. Online viewers will see a concatenated video of the orbital configuration of the system during each of the observations, including any potential PPO events that we found to occur during those windows. A still image of a PPO event during the observation on Oct 29, 2022 is included where the animation is not accessible. The top panel shows a bird's-eye view of the system with planet radii scaled up for better viewing. The distances and beam sizes are to scale, assuming a beam created with a 3.4m dish at 3.3~GHz (the maximum frequency observed during this particular session) from the surface of planet g aimed at planet e. The bottom panel shows the edge-on view with planet sizes scaled with distance, showing how much of the beam spills over the planet toward the direction of Earth in the negative z-direction. The red dashed lines in the illustrated beam is the inner angle blocked by the occulting planet, e. The blue dashed lines show the outer angle of the beam that would spill over the planet. The window for this event lasted roughly 95 minutes.}\label{fig:PPOs}
    \end{minipage}
\end{figure}

A modified version\footnote{\url{https://github.com/Tusay/NbodyGradient.jl}} of the \texttt{NbodyGradient} algorithm originally developed by \cite{NbodyGradient_Agol:2021:} was used to determine the orbital positions of the bodies in the TRAPPIST-1 system from August 23, 2015, 10:20:51.81 UT through the end date of the last observations for this project. Some of the modifications included adding radii of the bodies in the system, calculating an impact parameter to determine planet-planet occultations, and using smaller time steps (0.002 days down from 0.05 days) to ensure small timescale PPOs could be identified at any orbital phase. The modified code was tested to accurately replicate the results of the original \texttt{NbodyGradient}, which has been shown to predict mid-transit times based on transit timing variations for this particular system within microsecond agreement of similar N-body integrators, like \texttt{TTVFast} \citep{NbodyGradient_Agol:2021:}. At every time step, the algorithm calculates and records the positions of every body in the system. We assume these positions to be accurate outside of transit; the alignment of multiple bodies along the line of sight of the Earth was calculated during the times of these observations.

Uncertainties in the inclination, $i$, and semi-major axis, $a$, of the planets could still lead to alignments that disallow a PPO event. Figure \ref{fig:PPO_sketch_2} shows the geometry for which uncertainties in $a$ and $i$ have the maximum effect on observability of spillover radiation. The most extreme case was calculated for each pairs of planets during a potential PPO event. In most cases, the maximum dish diameter that could have been assumed and still created a PPO event at the frequencies observed was much larger than the dish diameter assumed.

The orbital configuration of the system during the time of observation based on the output of the \texttt{NbodyGradient} code was simulated as animated gifs. Figure \ref{fig:PPOs} shows a concatenation of the gifs, and the representative still frame captures the PPO event on October 29, 2022. The duration of the 7 PPO events during the 28 hours of observations ranged from 8.6 minutes to 99.4 minutes. Additional simulations of this system outside these windows indicate that over the course of just a few days many PPOs are likely to occur with a similarly wide range of durations, showing good agreement with previous studies, and offering encouraging results as a search strategy \citep{lugerPlanetPlanetOccultations2017, Reza_Ashtari:2023:}.

\section{Results}
\label{sec:results}

\medskip

The \texttt{NbeamAnalysis} pipeline does not output a definitive identifier of ETI, but instead maps each signal in the data onto a 2D parameter space that can be used as a guide for sorting and filtering the list of signals as candidates for follow-up. The two parameters are the DOT score, which scores similarity of signals in each beam, and the SNR-ratio, which measures the attenuation of signal power in each beam. The expectation is that true-positive, sky-localized signals will tend to have low DOT scores and high SNR-ratios, while local RFI will tend to have high DOT scores and low SNR-ratios. 

\begin{figure}
    \centering
    \begin{minipage}[hbt!]{\textwidth}
        \centering
        \includegraphics[width=0.9\textwidth]{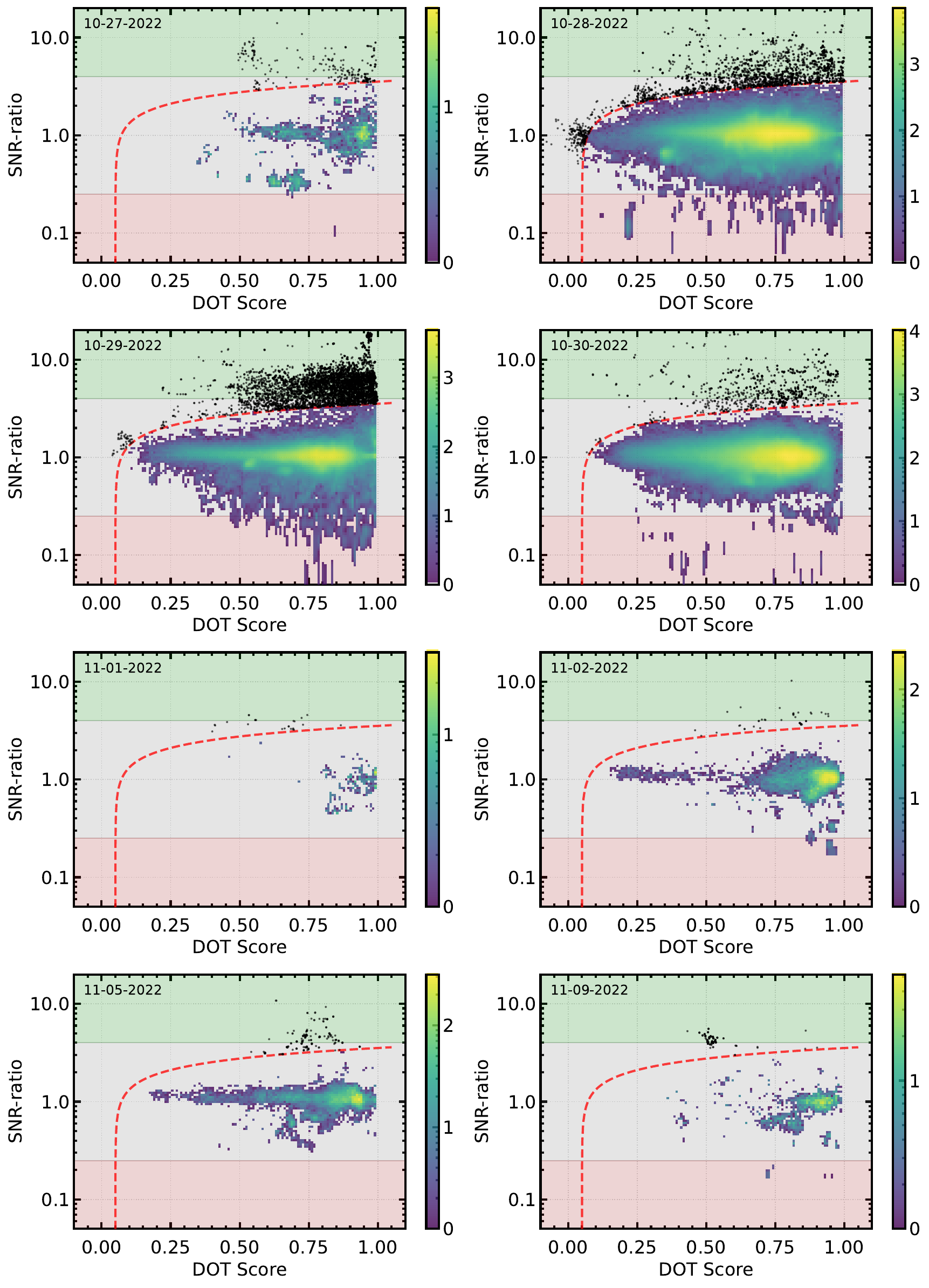}
        \caption{SNR-ratio vs. DOT scores for each day of the TRAPPIST-1 observations. Each observation was analyzed independently to account for temporal variability in the RFI environment. The red dashed line represents a conservative cutoff defined by an empirical power law designed to isolate the bulk of the RFI. RFI is expected to have high DOT scores and low SNR-ratios while true signals are expected to trend toward lower DOT scores and higher SNR-ratios. All signals above the cutoff are plotted as individual black points while signals captured below the cutoff are represented as a heatmap, with the colorbar showing the base 10 exponent of the number of signals represented in a given region.}
        \label{fig:SNRx}
    \end{minipage}
\end{figure}

\begin{figure}
    \centering
    \begin{minipage}[hbt!]{\textwidth}
        \centering
        \includegraphics[width=\textwidth]{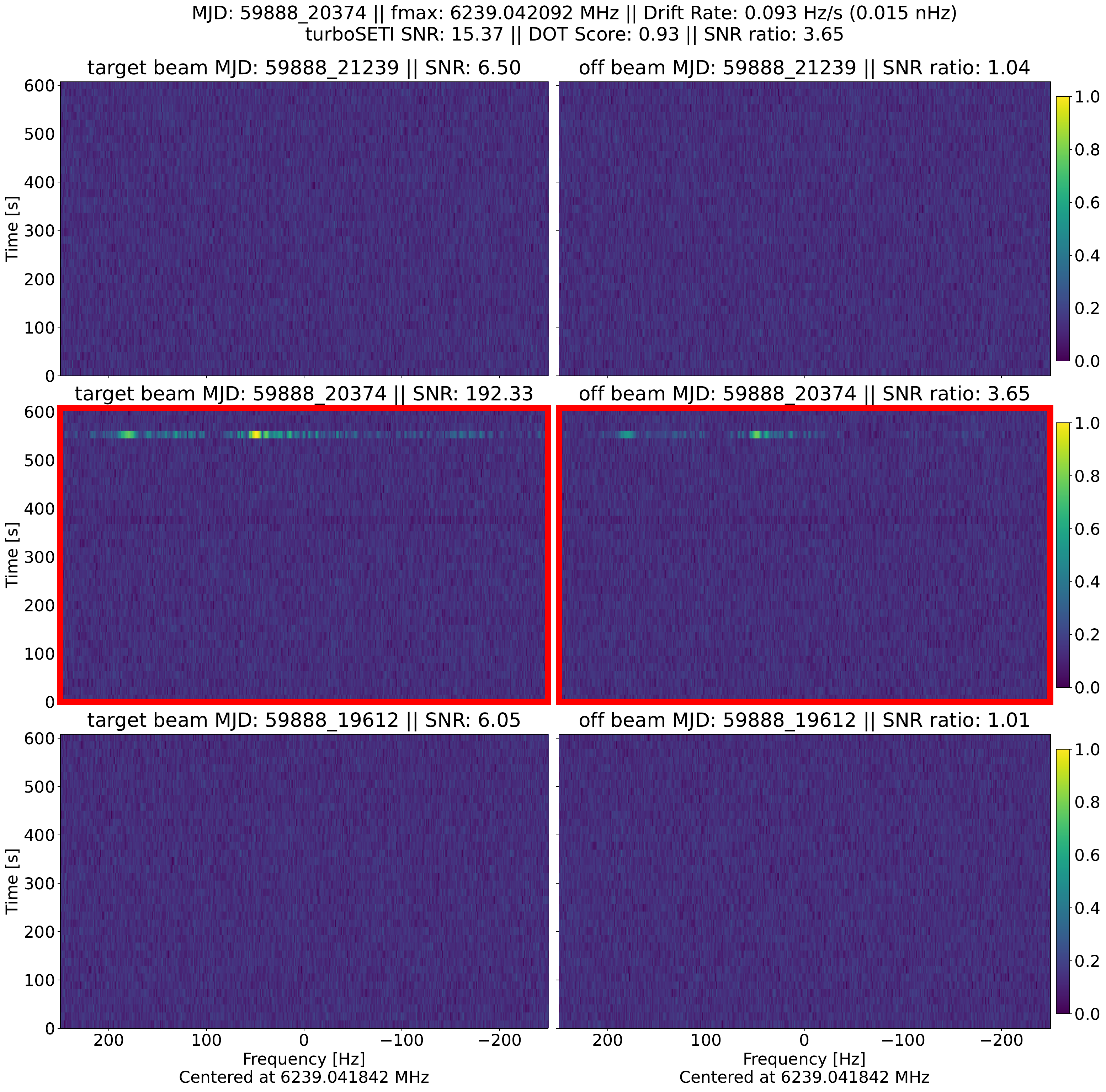}
        \caption{An example of some one-off ``blip" results. The red framed central panel is the 10-minute integration from the observations taken on Nov 05, 2022 in which the signal was found, showing reasonable attenuation but lacking the drifting narrowband morphology and persistence in time expected from a true signal in this kind of search. It appears to mimic other confirmed RFI in morphology, though at a distinct and isolated frequency. The observing time progresses from the bottom to the top of each panel. The bottom and top rows show adjacent integrations at the same frequency.} \label{fig:blips1}
    \end{minipage}
\end{figure}

\begin{figure}
    \centering
    \begin{minipage}[hbt!]{\textwidth}
        \centering
        \includegraphics[width=\textwidth]{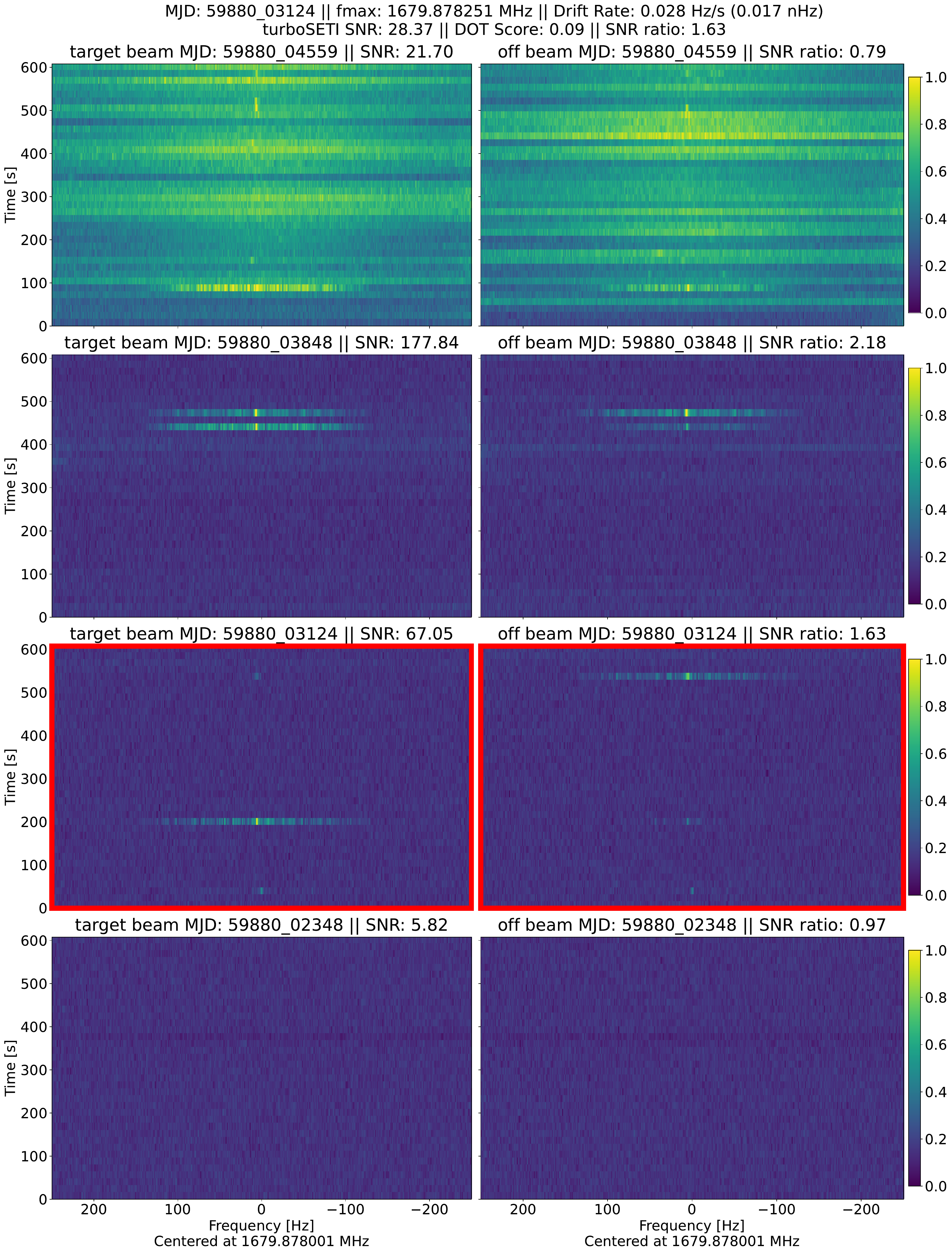}
        \caption{An example of unique and unexpected RFI in the primary panel framed in red. The observing time progresses from the bottom to the top of each panel. The other rows show adjacent integrations at the same frequency. The apparently alternating pings in the primary frame explain the low correlation score and spark curiosity. However, the low SNR-ratio, small drift rate, and bright features in later integrations all suggest that the source of the signal is likely not localized to the TRAPPIST-1 system.} \label{fig:weird}
    \end{minipage}
\end{figure}

The 2D parameter space for each observation of the TRAPPIST-1 data is plotted in Figure \ref{fig:SNRx}, showing that the vast majority of signals, over 99.8\% of the combined data, are identified as RFI. Each observation is treated separately to account for any temporal dependence in the RFI environment at the time of observation.

After filtering and scoring the list of signals, the pipeline also contains a plotting tool to show comparison dynamic spectra (waterfall plots) of each beam for a given signal, with the color scale representing the signal power and normalized to the power in the target beam. For practical considerations, the default configuration plots up to 500 of the most interesting candidate signals, though this number is adjustable as an input parameter. A nominal cutoff, defined in \S \ref{Mars}, is first used to identify obvious RFI with high DOT scores and low SNR-ratios, and if still more than 500 signals remain, the default plotting scheme will plot up to the lowest 500 scoring candidates above the default attenuation value. 

In the case of the TRAPPIST-1 observations specifically, 11127 candidate signals were plotted for visual inspection, of which 1627 were found to be within the PPO windows. We do not see a notable difference in the number of signals detected during PPO windows, and none of the signals indicated a definitive technosignature detection of non-human origin, though a handful warranted extra consideration. 

Figure \ref{fig:blips1} shows an example of a one-off ``blip" occurring over a timescale less than the 16-second time bins in the recorded data. Figure \ref{fig:weird} shows the most enigmatic example of various signals with an unexpected morphology. Many of these blips and morphologically interesting signals recovered by the pipeline were easily discounted as RFI when looked at more carefully, but a small subset, like the one shown in Figure \ref{fig:blips1} held too little information and dissimilarity to other signals in nearby frequency space. It may be worth searching for these signals in future observations. However, like the famed ``Wow!" signal \citep{Wow!_Kraus:1979:}, it seems imprudent to assign confidence to signals that do not persist and could easily be attributed to an increasingly crowded and complicated RFI environment. 

From this null result, we can calculate an upper limit on the signals that we were sensitive to in this search. The narrowband form of the radiometer equation from \citet[e.g., ][]{Enriquez_2017ApJ...849..104E} provides the spectral flux density $S_{min,narrow}$ in terms of the SNR, System Effective Flux Density (SEFD) of the instrument, the transmitter bandwidth $\Delta \nu_t$, the channel width $\Delta \nu$, the number of polarizations $n_{pol}$, and the total integration time per scan, $\tau_{obs}$:

\vspace{-0.3cm}
\begin{equation}
    \label{eq:radiometer_narrow}
    S_{min,narrow} = (\mathrm{SNR})\frac{\mathrm{SEFD}}{\Delta \nu_t}\sqrt{\frac{\Delta \nu}{n_{pol}\tau_{obs}}}
\end{equation}

The majority of these values were chosen when determining the data dimensions and survey design, however the SEFD must be calculated separately for each tuning and observing day as a measure of the sensitivity of the ATA under those particular conditions. 

We follow the same SEFD derivation procedure as \citet{sheikh2024characterization}. Briefly, each observing session began with a 10~min calibrator scan of a flux calibrator --- one of 3C286, 3C48, 3C295, and 3C84\footnote{3C84 is actually suitable for gain calibration, but not generally for flux calibration. In this case, for the first observing session only, 3C84 was accidentally used as the calibrator. In Summer 2023, we derived a set of coefficients for flux calibration with the ATA --- these values were used to use 3C84 as a pseudo-calibrator for this dataset.}. We used the measured gain values from the calibrator scan and the expected flux values for each calibrator \citep[generally from][except in the case of 3C84]{perley2017accurate}, to determine the SEFD of each antenna-polarization combination, and then average these values to get the SEFD of the beamformer as a whole.

We thus calculate $S_{min,narrow}$ for each session-tuning combination using SNR = 10, $\Delta \nu_t$ = $\Delta \nu$ = 1~Hz, $n_{pol}$ = 2, $\tau_{obs}$ = 10~min, and the SEFD for that session-tuning combination. We can then convert this spectral flux density into a required transmitter Equivalent Isotropic Radiated Power (EIRP) by using the following equation:

\vspace{-0.2cm}
\begin{equation}
    \label{eq:eirp}
    \mathrm{EIRP}_{min} = 4 \pi d^2 S_{min} \Delta \nu_t
\end{equation}

and plugging in $d$ = 12.5 parsecs, we find a range of potential EIRPs that we might be sensitive to, as reported in Table \ref{tab:observations}.

\section{Discussion \& Conclusion}
\label{sec:discussion}

\medskip

TRAPPIST-1 is inarguably an excellent laboratory for high precision multi-planet transiting exoplanet science. And by extension, it is perfect for refining targeted technosignature search strategies for leaked radio emission. The observational data presented here provide a useful reference for this system and the current RFI environment. The pipeline developed to search for and filter interesting signals in beamformed data at the ATA performs as well as current single dish search strategies, with similar limitations. However, long stares at every nearby multi-planet edge-on system is impractical. 

PPO events comprised roughly 21\% of the total observation duration in this experiment. And, while the TRAPPIST-1 system is one of the most favorable systems for PPOs, having 7 closely aligned known transiting planets, most other planetary systems where PPOs could plausibly occur are likely to have far fewer PPOs, leading to a smaller percentage of total time when it would be optimal to observe. Therefore, observing during predicted PPO events would significantly enhance the efficiency of targeted searches for leaked transmission in these systems.  

The DSN transmits continuously as it tracks space probes, and frequently schedules tracking for several hours at a time. We cross-referenced recent scheduling from the Canberra station\footnote{\url{https://www.cdscc.nasa.gov/Pages/trackingtoday.html}} with a recent NASA audit of the DSN\footnote{\url{https://oig.nasa.gov/docs/IG-23-016.pdf}} that includes tracking hours for the most subscribed satellites to estimate how often the DSN is transmitting toward a particular target. Focusing just on Mars transmissions and adjusting the reported time by 80\% to account for target acquisition and other set up overhead, we estimate that the DSN transmits to Mars about 30-35\% of the time over a given period of months, typically for several hours at a time. 

The PPO event durations are significantly shorter than typical DSN transmissions and each PPO event found during these observations constitutes a unique pairing of planets. We treat each PPO event as an individual probability event within the scope of the entire observations and assume the duration of each PPO event does not significantly contribute to the overall probability of detecting an equivalently intermittent DSN. Therefore, we estimate the probability of having intercepted a transmission during one of these events to be roughly 33\% per PPO event. To have a significant chance of intercepting leaked emission from an intermittent transmission source similar to the current DSN, one would want to observe for a minimum of 3 PPO events for each pairing of planets. 

There are of course many caveats to this consideration, as it does not account for all possible factors that could affect the probability of detection from an equivalent ETI DSN. We cannot guess at the true nature of an alien communication system, so we offer this calculation as a first order approximation based entirely on the primary system in use by the only currently known species to communicate through deep space: the human-made DSN. We leave the inclusion and refinement of additional factors that may better inform search strategies along this vein to future research.

The TRAPPIST-1 system has many advantages that have led to the feasibility of determining and predicting PPO events, but the same process remains difficult and time-intensive for other systems. Future work will include adapting such techniques to other applicable systems, especially as higher-precision orbital data becomes available for similar, co-planar systems.

\begin{figure}
    \centering
    \begin{minipage}[hbt!]{\textwidth}
        \centering
        \includegraphics[width=\textwidth]{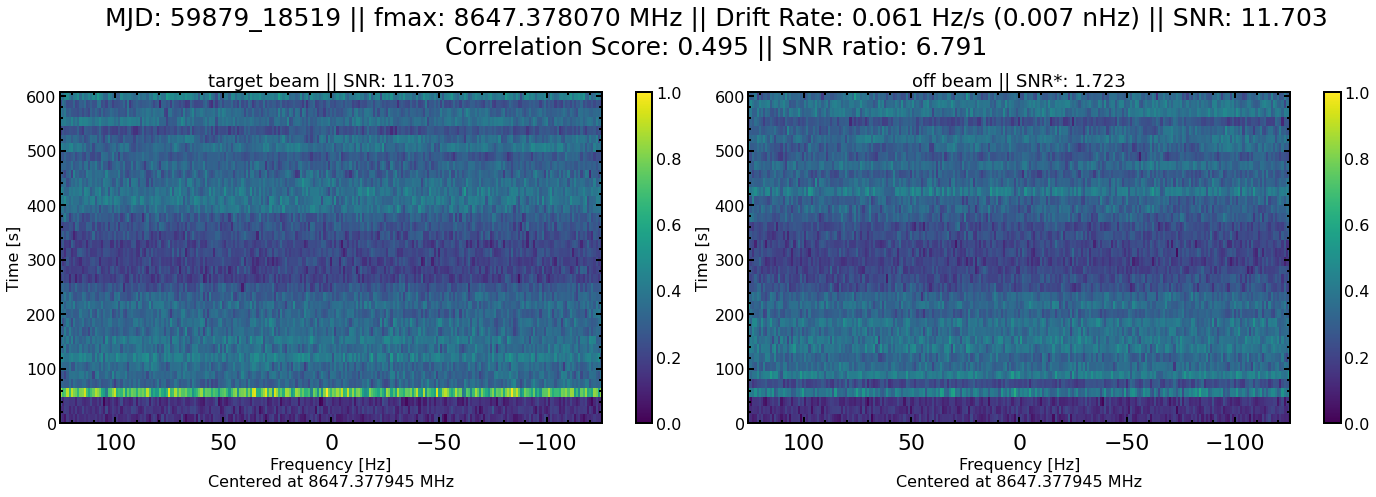}
    \end{minipage}
    \begin{minipage}[hbt!]{\textwidth}
        \centering
        \includegraphics[width=\textwidth]{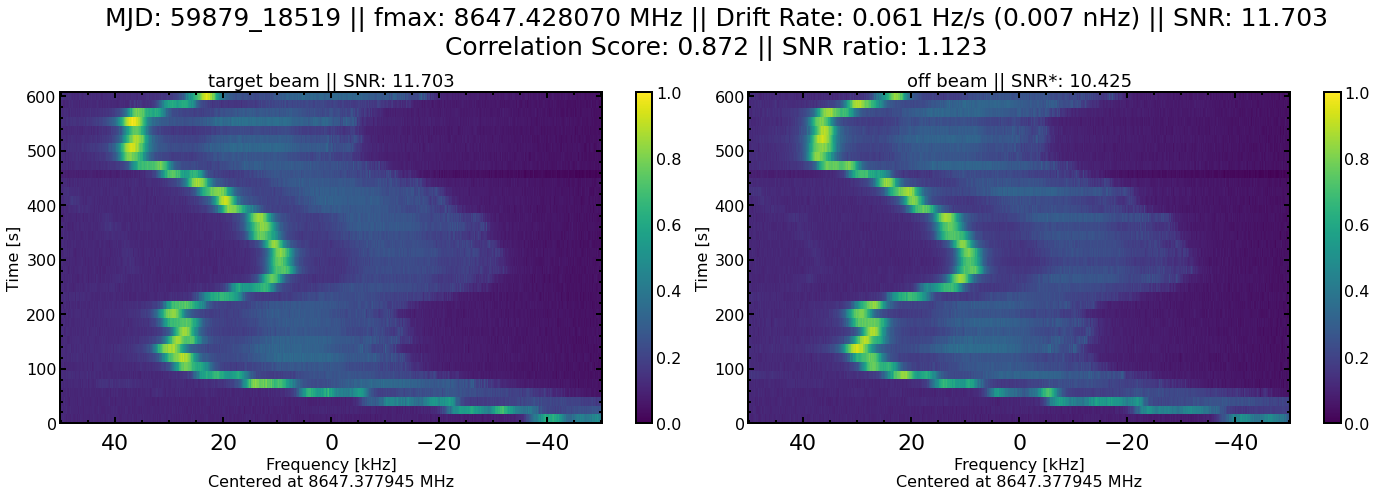}
        \caption{Original plot of a signal over the expected frequency span based on drift rate in the top row, and the same signal zoomed out in frequency in the bottom row. The true morphology of this broadband signal can be seen over a wider bandwidth and is reflected in the increase in correlation (DOT) score. This demonstrates how small slices of broadband signals are often misinterpreted by current de-Doppler algorithms as narrowband signals.}\label{fig:broadbandRFI2}
    \end{minipage}
\end{figure}

While great strides have been made to automate the process of detecting signals and filtering out obvious RFI in very large datasets with very high resolution like this, some challenges remain difficult to overcome. As mentioned in \S\ref{setigen} and shown in Figure \ref{fig:injection3}, low-power signals from a distant source overlapping with much stronger RFI at the same frequency may easily be missed with this and other similar search and filter algorithms. 

Another confounding problem is broadband RFI as illustrated in Figure \ref{fig:broadbandRFI2}. Slight fluctuations in power between beams often presents as different power signatures in narrowband channels over smaller scales than the true width of the broadband signal. In some cases, this may be erroneously interpreted as distinct narrowband signals by the search algorithm. This can also lead to low correlation (DOT) scores in the filtering pipeline. Although the signals are easily discerned as part of the same broadband feature when zoomed out to larger bandwidths, identifying and classifying these structures as broadband interference is a poorly defined problem and difficult to automate. 

Recent attempts at employing machine-learning techniques have seen some success in retrieving previously undetected signals in non-AI driven pipelines \citep{Ma:2023:, ML_Ma:2024:}. Machine learning algorithms have proven very effective as classifiers and identifying key features in image data, in certain instances. It is very possible that some adaptation of these techniques may be useful for mitigating these problems of broadband characterization and signals hidden behind RFI. 

The plotting software in the \texttt{NbeamAnalysis} pipeline offers the ability to stack integrations, which can help track persisting signals and provide additional information for identifying the nature of the signal. However, as seen in Figure \ref{fig:blips1}, many signals occur over shorter timescales than the 16s time binned data, which provides almost no information on its drift rate unless the signal pings again. Preserving the sampling at shorter timescales without binning may help, but this would significantly increase an already dense data volume. 

Similarly, along the orthogonal axis, this project attempted to apply frequency binning (fscrunching) as a potential solution to help recover power lost at higher drift rates. But, when applying the fscrunching technique, the de-Doppler algorithm as currently implemented with \texttt{turboSETI} is run anew on the frequency binned data. Therefore, the algorithm will sometimes trigger on and assign a higher drift rate to the same signal at wider frequency ranges. So, many of the hits, particularly at higher drift rates, are seen to be duplicates of the same signal upon further inspection. Attempting to recover weak signals at high drift rates may be better handled through direct integration of frequency binning into the de-Doppler algorithm from the start, if not an entirely new process to recover signals at high drift rates without substantial loss of power in the signal. 

The limits of the DOT algorithm on ATA data need to be better characterized, including determining a more quantitative cut off value for future searches. This should be paired with a more robust injection-recovery analysis. In this project, a simple beam-localization procedure was used: the on-beam was always at the phase-center of the array, and the off-beam was always placed a constant separation, of 9 arcminutes, from the phase center within the primary beam. However, the beam isolation would be improved by using a variable off-beam location, adjusted to fall within a ``null'' of the on-beam's pattern. 
Work is ongoing to incorporate a model of the beamshape that would also allow for a more precise determination of the attenuation in the off-beam (not just the threshold value of 4 used in this work), reducing the number of false positive signals that need to be manually vetted. 

The analysis of the observations presented here demonstrates that precise characterization of ideal systems, like TRAPPIST-1, enabling orbital dynamical modeling and prediction of PPO events offer practical application for leaked emission searches. This provides SETI a powerful new observational tool and search strategy. As signal detection and RFI mitigation pipelines improve, the inclusion of PPOs to provide narrow search windows may make it more feasible to increase time resolution and sensitivity at higher drift rates. 

The minimum detectable transmitter power by the ATA during these observations, listed in Table \ref{tab:observations}, show that much greater sensitivity is needed to detect transmissions with power on the order of the DSN, which typically operates at tens of kiloWatts of directed transmission, and $\sim$10 GW EIRP\citep{DSN_power_Derrick:2023:}. It is technically feasible that an Arecibo-like transmitter---operating at 20 TW EIRP---would be detectible by the ATA at low to moderate drift rates, but such powerful transmitters are generally seen as overkill, especially in the compact TRAPPIST-1 system, and not regularly used as part of the DSN for interplanetary communication. However, future observatories, such as the Square Kilometer Array may provide the sensitivity required to detect levels of leaked emission comparable to the DSN \citep{SKA1_Sensitivity_Braun:2019:}. Detection of leaked radio emission from nearby planetary systems may be on the horizon.

\vspace{-0.3cm}
\section*{Acknowledgments}

We acknowledge Daniel Estévez for assistance with the DSN transmission time calculation and Megan Li for assistance with the drift rates expected from the TRAPPIST-1 system. We would like to acknowledge the student and chaperone observers from the 2023 SETI Institute REU program who contributed the Mars downlink data, whose participation was enabled by NSF award No. 2051007: Alex Medina, Alyssa Jankowski, Ella Hort, Fatima Saccoh, Jasmine Freeman, Maddy Korkeakoski, Madigan Rumley, Megan Pirecki, Nia Butler, Reed Spurling, Russell Mapaye, Samantha Hemmelgarn, Steven Gracy, and Alex Jackson [students] and Rosalba Bonaccorsi, Rob French, Debra Stopp, and John Bowman [chaperones].

N.T. acknowledges that this material is based upon work supported by the National Science Foundation Graduate Research Fellowship Program under Grant No. DGE1255832. 

S.Z.S. acknowledges that this material is based upon work supported by the National Science Foundation MPS-Ascend Postdoctoral Research Fellowship under Grant No. 2138147.

The authors acknowledge the Penn State Extraterrestrial Intelligence Center and the Penn State Center for Exoplanets and Habitable Worlds, which are supported by the Pennsylvania State University, and the Eberly College of Science. 



\facility{ATA}
\software{turboSETI (Enriquez et al. 2017), NbodyGradient (Agol et al. 2021)}

\bibliography{paper}{}
\bibliographystyle{aasjournal}

\end{document}